\documentclass[aoas,preprint]{imsart}

\input{package_newcomment}
\usepackage{multirow}
\usepackage{algpseudocode,algorithm}
\usepackage{lipsum}

\begin{document}

\begin{frontmatter}
\title{Discovering Political Topics in Facebook Discussion threads with Graph Contextualization}
\runtitle{Spectral Contextualization}

\begin{aug}
\author{\fnms{Yilin} \snm{Zhang}\thanksref{t1}\ead[label=e1]{yilin.zhang@wisc.edu}},
\author{\fnms{Marie} \snm{Poux-Berthe}\thanksref{t2}\ead[label=e5]{m.poux-berthe@live.fr}},
\author{\fnms{Chris} \snm{Wells}\ead[label=e3]{cfwells@wisc.edu}},
\author{\fnms{Karolina} \snm{Koc-Michalska}\thanksref{t2}\ead[label=e4]{kkocmichalska@audencia.com}},
\and
\author{\fnms{Karl} \snm{Rohe}\thanksref{t1}\ead[label=e2]{karlrohe@stat.wisc.edu}}

\thankstext{t1}{The authors gratefully acknowledge support from NSF grant DMS-1612456 and ARO grant W911NF-15-1-0423.}
\thankstext{t2}{The authors gratefully acknowledge support from Audencia Foundation Research grant.}
\runauthor{F. Author et al.}
\affiliation{University of Wisconsin-Madison and Audencia Business School}

\address{
Yilin Zhang, Karl Rohe\\
Department of Statistics\\
University of Wisconsin Madison\\
1300 University Ave\\
Madison, WI 53706\\
USA\\
\printead{e1}\\
\phantom{E-mail:\ }\printead*{e2}}

\address{
Karolina Koc-Michalska, Marie Poux-Berthe\\
Audencia Business School\\
Communication and Culture Department\\
1 Rue Marivaux\\
44003 Nantes\\
France\\
\printead{e5}\\
\phantom{E-mail:\ }\printead*{e4}}
\end{aug}

\address{
	Chris Wells\\
	School of Journalism and Mass Communication\\
	University of Wisconsin Madison\\
	5115 Vilas Hall\\ 
	821 University Avenue\\
	Madison, WI 53706\\
	USA\\
	\printead{e3}}
\begin{abstract}
We propose a graph contextualization method,  \texttt{pairGraphText}, to study political engagement on Facebook during the 2012 French presidential election.  It is a spectral algorithm that contextualizes graph data with text data for online discussion thread.  In particular, we examine the Facebook posts of the eight leading candidates and the comments beneath these posts.  We find evidence of both (i) candidate-centered structure, where citizens primarily comment on the wall of one candidate and (ii) issue-centered structure (i.e. on political topics), where citizens' attention and expression is primarily directed towards a specific set of issues (e.g. economics, immigration, etc).  To identify issue-centered structure, we develop \texttt{pairGraphText}, to analyze a network with high-dimensional features on the interactions (i.e. text). This technique scales to hundreds of thousands of nodes and thousands of unique words. In the Facebook data, spectral clustering without the contextualizing text information finds a mixture of (i) candidate and (ii) issue clusters.  The contextualized information with text data helps to separate these two structures. 
We conclude by showing that the novel methodology is consistent under a statistical model. 
\end{abstract}


\end{frontmatter}

\section{Introduction}

Social networking sites (SNSs) such as Facebook and Twitter now make up a major part of Internet communications (\cite{ellison2007social}, \cite{kaplan2010users}), including political communication.  By providing platforms for citizens to publicly communicate with each other and with politicians, SNSs may increase the accessibility of candidates and political dialog \citep{wellman2001does} and motivate political engagement within the public (\cite{williams2013social}, \cite{williams2009explaining}, \cite{hebshi2011online}, \cite{kushin2009getting}).  They also appear to facilitate the spread of false or offensive information, and a variety of forms of actors to reach micro-targeted publics with a high degree of efficiency \citep{kreiss2017technology}. Since the 2008 US election particularly \citep{wattal2010web}, SNSs have been playing a significant role in advertising and interactions during the presidential elections.

Drawing meaning from the massive text corpora of political discussion threads on SNSs has been a major project of scholars working in text mining (\cite{pang2008opinion}, \cite{stieglitz2013social}, \cite{grimmer2013text}) and sentiment analysis in recent years.  One popular text mining approach is the probabilistic topic models based on latent Dirichlet allocation (LDA) (\cite{blei2003latent}, \cite{blei2012probabilistic}, \cite{chang2009relational}), which have been extensively used in social science \citep{ramage2009comprehensive}.  Sentiment analysis is another approach to analyze text.  It focuses on understanding emotions in the text.  \cite{wang2012system} provides a system for real-time sentiment analysis on Twitter during the 2012 US election.  For instance, using sentiment analysis and regression, \cite{stieglitz2012political} finds that political tweets on Twitter that contain stronger emotions receive more public interactions.  There are also studies of how political sentiment on SNSs reflect the offline political landscape \citep{tumasjan2011election}, and how it can affect political elections \citep{choy2011sentiment}.

Apart from the topic or sentiment information, patterns of political discussion on SNSs are also of great theoretical and empirical interests to scholars of communication and political science. Such platforms have long been heralded for their potential to foster a ``public sphere" in which ordinary citizens can recognize one another and hear reasons both for and against their own points of view (\cite{papacharissi2002virtual}). More recent analyses of online political discourse are less optimistic, identifying instead vitriol, ``trolling", and larger patterns of partisan polarization. As a result, a great deal of research investigates the extent to which online actors are connected to political opponents (\cite{adamic2005political}, \cite{colleoni2014echo}, \cite{bakshy2015exposure}) 

Another approach to understand structure of political discussions is social network analysis, which aims to identify influential political actors and communities in the discussions \citep{stieglitz2012political} and to study properties of the communities (\cite{robertson2010off}) \cite{gonzalez2010structure}).  One popular community detection approach is spectral clustering \citep{von2007tutorial}, which is fast, easy to implement, and consistent in block models for network (\cite{holland1983stochastic}, \cite{airoldi2008mixed}, \cite{qin2013regularized}). 

In this paper, we combine text mining and community detection to investigate the multiple dimensions of citizens' interactions with political content coming from political actors.  In our data, which come from the 2012 French election, citizens commented on presidential candidate's Facebook posts.  This creates a communication network between two types of units: (i) citizens and (ii) candidate-posts, as the eight presidential campaigns each has posts on Facebook, and citizens comment on the posts. This paper studies the structure of the resulting discussion threads.

The activities of the citizens are characterized by (i) which of the candidate-posts they comment on and (ii) the text of their comments.  We are interested in two broad types of patterns in these activities: (i) candidate-centered structure, where citizens primarily comment on the wall of one candidate; and (ii) issue-centered structure, in which citizens' attention and expression is directed towards a specific set of issues (e.g. economics, immigration, etc).  To search for such patterns, we cluster the citizens based on their activities.  In each cluster, we examine whether the activities of the citizens focus on particular candidates (i.e. candidate-centered)(Section \ref{sec:fav_ratio}) or whether the activities focus on certain political issues (i.e. issue-centered)(Section \ref{issue_structure}). This distinction reflects the possibility that the Facebook conversation might be organized more along lines of partisanship (candidate-centered), as opposed to matters of concern to ``issue publics" (issue-centered) (\cite{kim2009issue}).  

There has been significant progress on both topic modeling for text \citep{blei2012probabilistic} and community detection for social networks (\cite{airoldi2008mixed}). Recently, there has been significant interest in clustering networks for which we have additional information on the citizens in networks (\cite{chang2010hierarchical}; \cite{binkiewicz2017covariate}).  In this paper, we extend these ideas to the setting of discussion threads.  Our network is two-way or bi-partite, in which the two types of units, citizens and candidate-posts, are linked by commenting in a discussion thread.  Below, we refer to the links showing which citizens commented on which candidate-posts as the \textbf{network} or the \textbf{graph}.  We refer to both the text in candidate-posts and the text in citizen-comments as the \textbf{text}.  The duality between citizens and candidate-posts also appears in the text; candidates say things differently from citizens. 

A key difficulty in analyzing this process, and the key methodological innovation of this paper, is to combine these disparate sources of data, the graph information and the two types of text information (citizen-words and thread-words), in a meaningful way.  We develop a graph contextualization technique, \texttt{pairGraphText}, to leverage high dimensional node covariates into spectral clustering.  We extend and specialize the techniques of \cite{binkiewicz2017covariate} to deal with both (i) the asymmetrical nature of the network between citizens and candidate-posts, and (ii) the high dimensional and sparse nature of the text.  With noticeable themes, four sub-populations and four sub-groups of the candidate-posts are uncovered by our method.  We interpret the clusters by a word-content strategy: For each cluster, we (i) identify keywords, and then (ii) read through central conversations containing the keywords. 

Our graph contextualization method, \texttt{pairGraphText}, is adaptable to symmetric or directed graphs, unipartite or bipartite, assortative or dis-assortative, weight or unweighted.  It scales to hundreds of thousands of nodes and thousands of covariates (e.g. words).  \texttt{pairGraphText} uses a sparsity penalty to select the key covariates that align with the graph.  After combining the covariates with the graph, we use spectral clustering to compute a partition of the nodes.  Finally, we provide diagnostics to identify key covariates to interpret the different clusters.  Theorem $\ref{thm:upper}$ shows that our method is consistent under the Node-Contextualized Stochastic co-Blockmodel.  Section \ref{issue_structure} uses \texttt{pairGraphText} to identify the issue centered structure in the Facebook discussion threads.  In Section \ref{sec: compare}, we compare \texttt{pairGraphText} to a state-of-the-art topic modeling method, relational topic model (RTM) \citep{chang2009relational}, by both the Facebook discussion threads and simulations.  We show that RTM focuses more on the text data, while \texttt{pairGraphText} focuses more on the graph data.

This paper is organized as follows.  In Section $\ref{background}$, we briefly describe the 2012 French presidential election, the discussion threads on Facebook, and the result of regularized spectral clustering without any contextualizing information.  In Section $\ref{spectral_contextualization}$, we introduce the graph contextualization technique, \texttt{pairGraphText}, which leverages node covariates in spectral clustering.  In Section $\ref{issue_structure}$, we identify the issue-centered structure of the discussion threads using \texttt{pairGraphText}.  The statistical consistency of our method is provided under the Node Contextualized Stochastic co-Blockmodel in Section $\ref{sec:consistency}$.   In Section $\ref{sec: compare}$, we discuss different choices for weights of words, and we compare \texttt{pairGraphText} with a state-of-the-art topic modeling method.  Section $\ref{discussion}$ concludes with a discussion of our method. 

\section{Background and key summaries of the data}\label{background}

France's presidential elections proceed in two stages. On April 22 2012, the first round of voting narrowed the field of candidates from ten to two; the second round, between Fran\c cois Hollande and Nicolas Sarkozy, took place on May 6. In these analyses, we focus on the eight candidates who received at least $1\%$ of the votes in the 1st round of the election.  These eight candidates--Fran\c cois Hollande, Nicolas Sarkozy, Marine Le Pen, Jean-Luc M\'elenchon, Fran\c cois Bayrou, Eva Joly, Nicolas Dupont-Aignan, and Philippe Poutou--made a total of 3239 posts on Facebook.  In response, 92,226 Facebook users, which we call citizens, made 594,685 comments on the candidate-posts.\footnote{The data was gathered by \url{sotrender.com}.  They collect all posts from the official Facebook profiles of the top eight candidates and all the comments beneath them.  Citizens who commented on the candidate-posts are distinguished by identification numbers, which are corresponding to the urls of their Facebook profiles.  \url{sotrender.com} does not control for citizens being human users (non-bots) or being unique users (e.g. without establishing artificial accounts in order to comment on candidate-posts). }  

There are two main structures that we aim to detect and study in the conversation: (i) candidate-centered structure, where citizens primarily comment on the wall of one candidate; and (ii) issue-centered structure, in which citizens' attention and expression is directed towards a specific set of issues (e.g. economics, immigration, etc).

\subsection{The communication network}

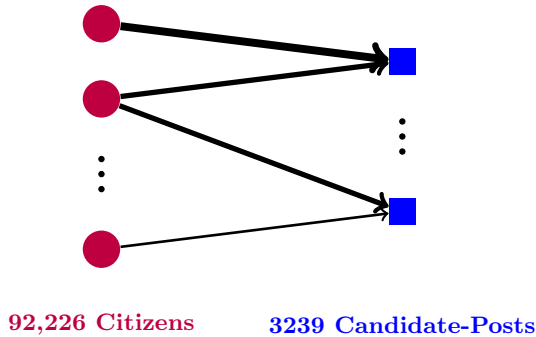
\begin{figure}[H]
\begin{center}
	\begin{tikzpicture}

\node[circle,fill=purple,inner sep=5pt,minimum size=3pt] (c1) at (-2,3) {};
\node[circle,fill=purple,inner sep=5pt,minimum size=3pt] (c2) at (-2,2) {};
\node[circle,fill=purple,inner sep=5pt,minimum size=3pt] (c3) at (-2,0) {};

\node[rectangle,fill=blue,inner sep=5pt,minimum size=3pt] (p1) at (2,2.5) {};
\node[rectangle,fill=blue,inner sep=5pt,minimum size=3pt] (p2) at (2,0.5) {};

\filldraw
(-2,0.8) circle(1pt)
(-2,1) circle(1pt)
(-2,1.2) circle(1pt)
(2,1.3) circle(1pt)
(2,1.5) circle(1pt)
(2,1.7) circle(1pt);

\draw (-2,-1) node[purple, font = \bf]{92,226 Citizens};
\draw (2,-1) node[blue, font = \bf]{3239 Candidate-Posts};


\draw[->, line width = 3pt](c1) -- (p1);
\draw[->, line width = 2pt](c2) -- (p1);
\draw[->, line width = 2pt](c2) -- (p2);
\draw[->, line width = 1pt](c3) -- (p2);

	\end{tikzpicture}
\end{center}
\caption{\textbf{The Communication Network} is a bi-partite graph between citizens and candidate-posts. Each edge weight corresponds to the number of times that a citizen comments on a candidate-post. }
\label{fig:network_plot}
\end{figure}

\begin{figure}[H]\centering 
	\subfigure[][]{\label{fig:plot_citidegree}
		\includegraphics[height=0.35\columnwidth]{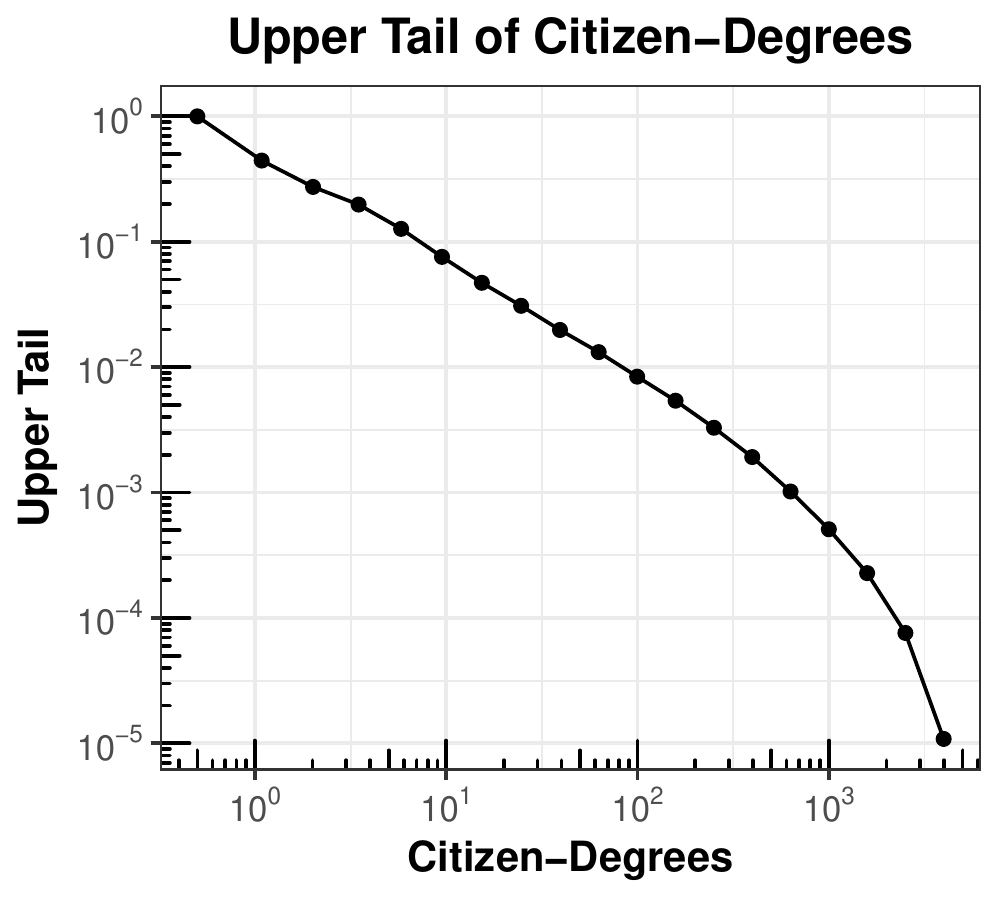}
	}
	\subfigure[][]{\label{fig:plot_postdegree}
		\includegraphics[height=0.35\columnwidth]{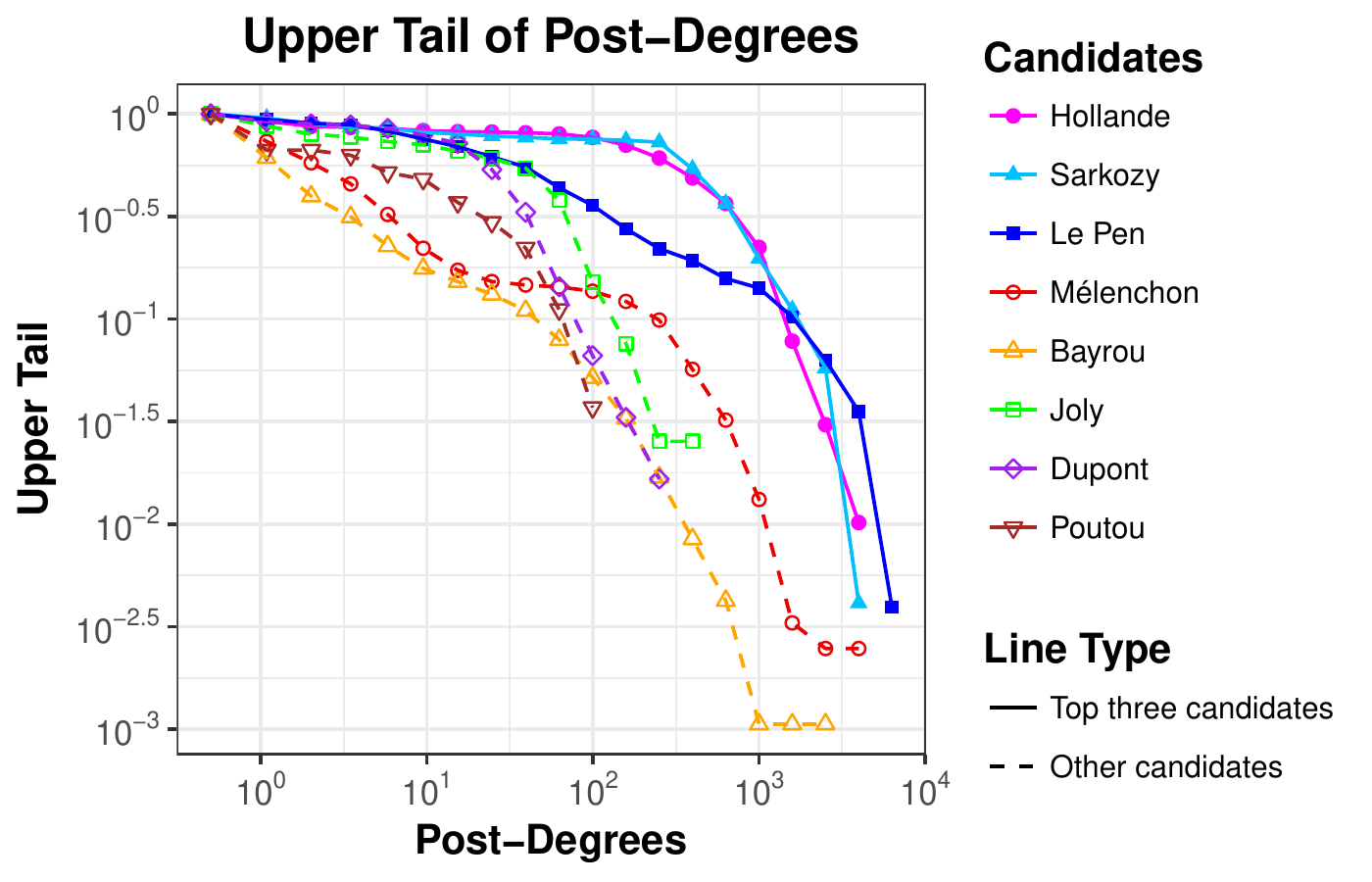}
	}
	\caption{\textbf{Upper Tail of Degrees.} Figure (a) shows the upper tail of citizen-degrees.  $90\%$ of the citizens write fewer than 10 comments, a small number of citizens write thousands of comments.  Figure (b) shows the upper tail of post-degrees by candidate.  The top three candidates: Hollande, Sarkozy, and Le Pen (on right), have the largest degrees.}
	\label{fig:plot_degree}
\end{figure}

To study the structure of the conversations, we construct a weighted bi-partite network between citizens and candidate-posts (see Figure $\ref{fig:network_plot} $) from the discussion threads.  
A citizen is linked to a candidate-post if and only if the citizen comments on the candidate-post.  The weight of this link is the number of times the citizen comments on the candidate-post.  To represent this network, we construct the weighted adjacency matrix $A \in \mathbb{R}^{92,226 \times 3239}$ with
\begin{equation}\label{eq:adjMat}
A_{ij} = \# \text{ of times of citizen $i$ comments on candidate-post $j$.}
\end{equation}
Denote the degree of a citizen $i$, $d_i = \sum_{j}A_{ij}$, as the number of comments by citizen $i$.  Denote the degree of a candidate-post $j$, $d_j = \sum_{i}A_{ij}$, as the number of comments underneath the candidate-post.  
Figure \ref{fig:plot_citidegree} shows the proportion of citizens who have at least $d$ comments, as a function of $d$.  Figure \ref{fig:plot_postdegree} gives the same result for the post-degrees. 

\subsection{Citizens' \text{attention-ratio} towards candidates}\label{sec:fav_ratio}

Let $\zeta_{ij}$ be the number of times that citizen $i$ comments under candidate $j$'s wall.  For each citizen $i$, we denote their \texttt{attention-ratio} as $$\text{AttentionRatio}(i) = \frac{\max\limits_{\ell}\zeta_{i\ell}}{d_i}.$$
When the attention-ratio is one, it indicates the citizen only comment on one candidate-wall, while smaller attention-ratio indicates the citizen comments across different candidate-walls. We say that citizen $i$ focuses on candidate $j$ if $\zeta_{ij} \geq \zeta_{i\ell}$ for any candidate $\ell$.  The citizens that have tied favorites are randomly assigned to one of their favorite candidates.  Then, the citizens are naturally partitioned into eight clusters based on the candidates they focus on.  Figure \ref{fig:favratio} shows the histogram of attention-ratio for all citizens with $d_i \ge 10$.  Most of the mass of this histogram is close to one, indicating that most citizens primarily comment on one candidate-wall. This gives the first impression of candidate-centered structure.

\myfig{1}{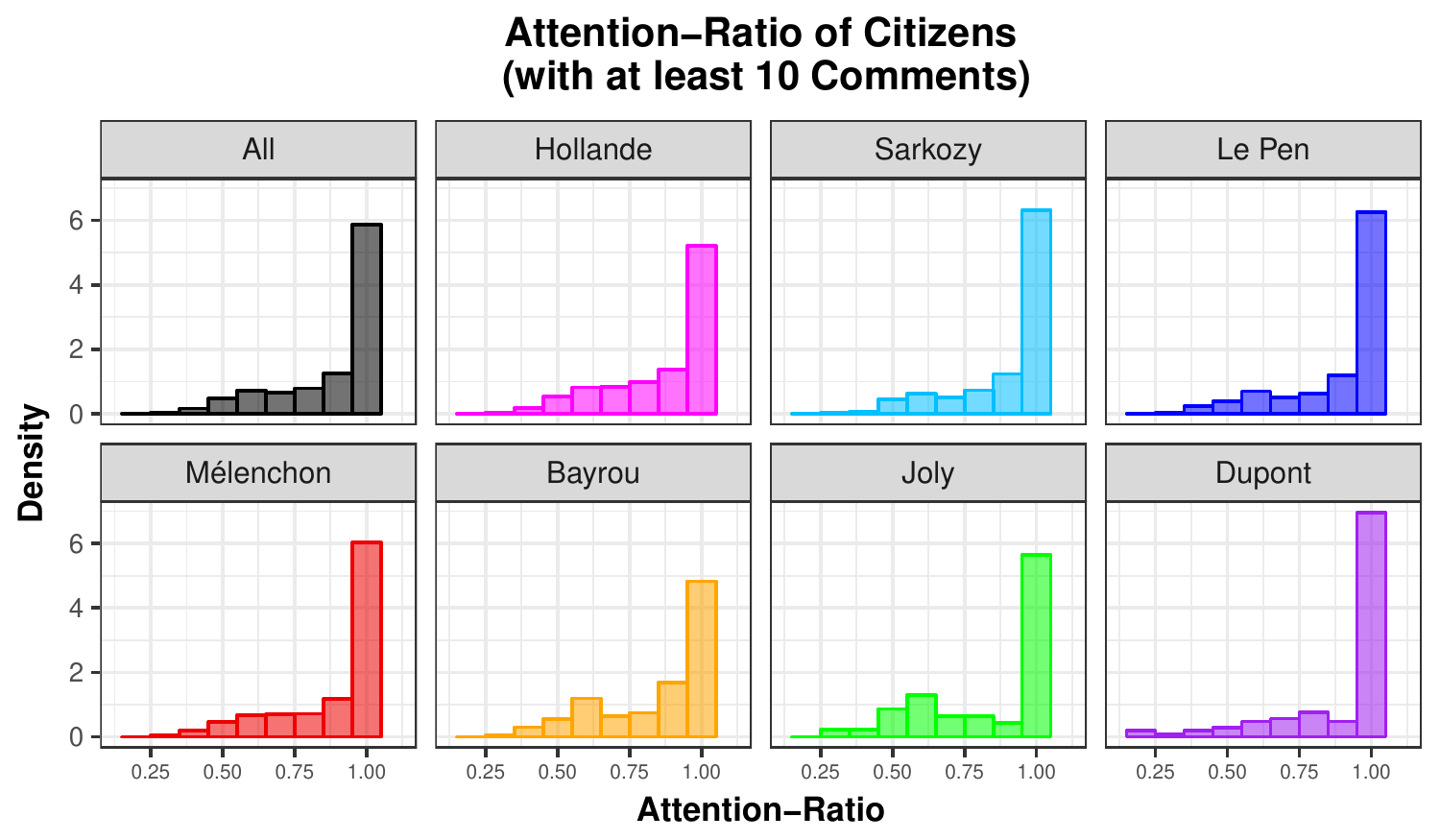}{\textbf{Distribution of Citizens' Attention-Ratio.} In this figure, we focus on citizens who have at least 10 comments. The first plot displays the histogram of attention-ratio for all citizens. The rest eight plots are for the eight citizen-clusters based on the candidates they focus on. We don't display the citizens who focus on Poutou, because he attracts very few comments.}{fig:favratio}

\begin{figure}[H]\centering 
\subfigure[][]{\label{fig:favratio_balloon}
\includegraphics[width = 0.49\columnwidth]{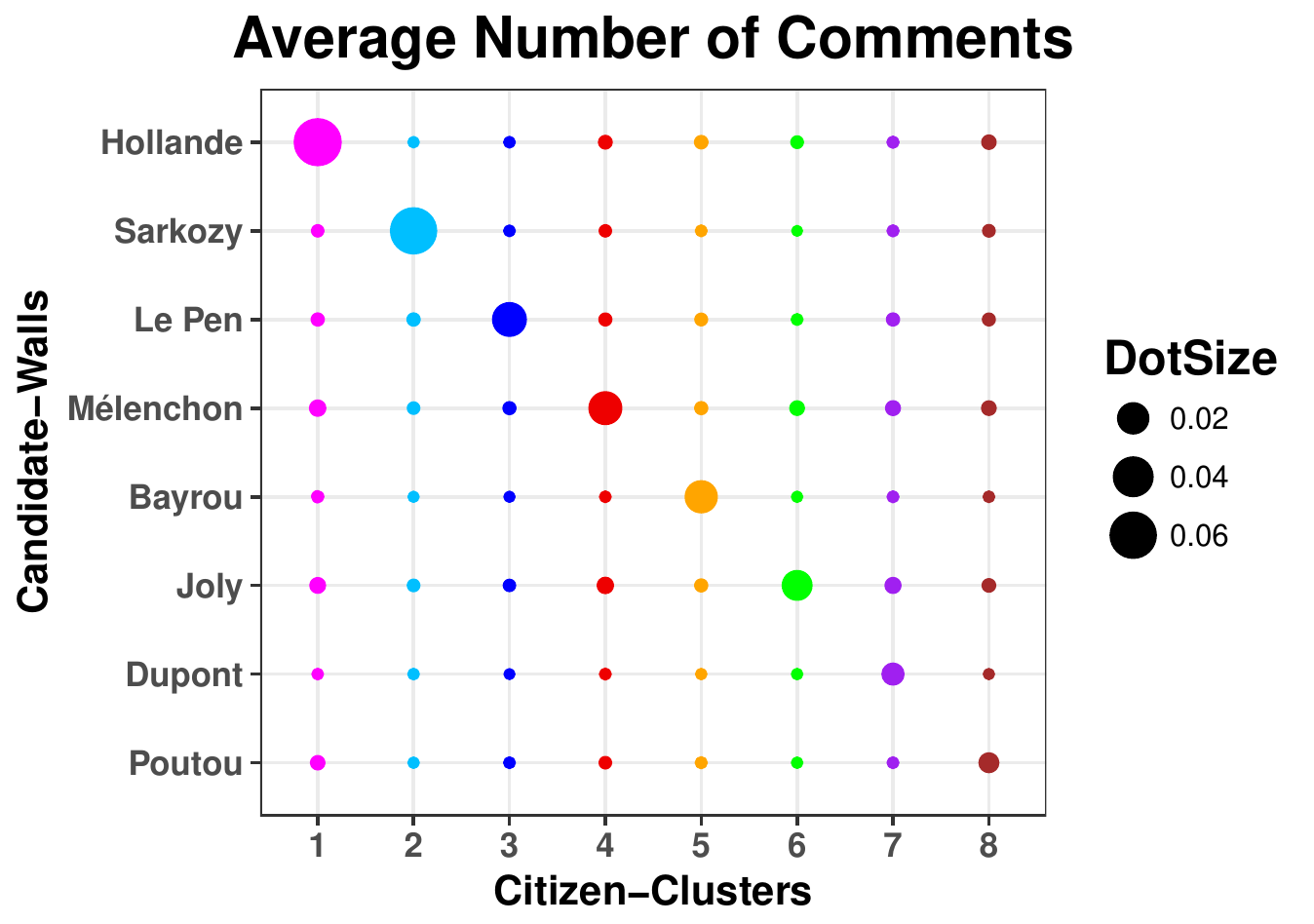}
}
\subfigure[][]{\label{fig:favratio_citisizes}
\includegraphics[width = 0.46\columnwidth]{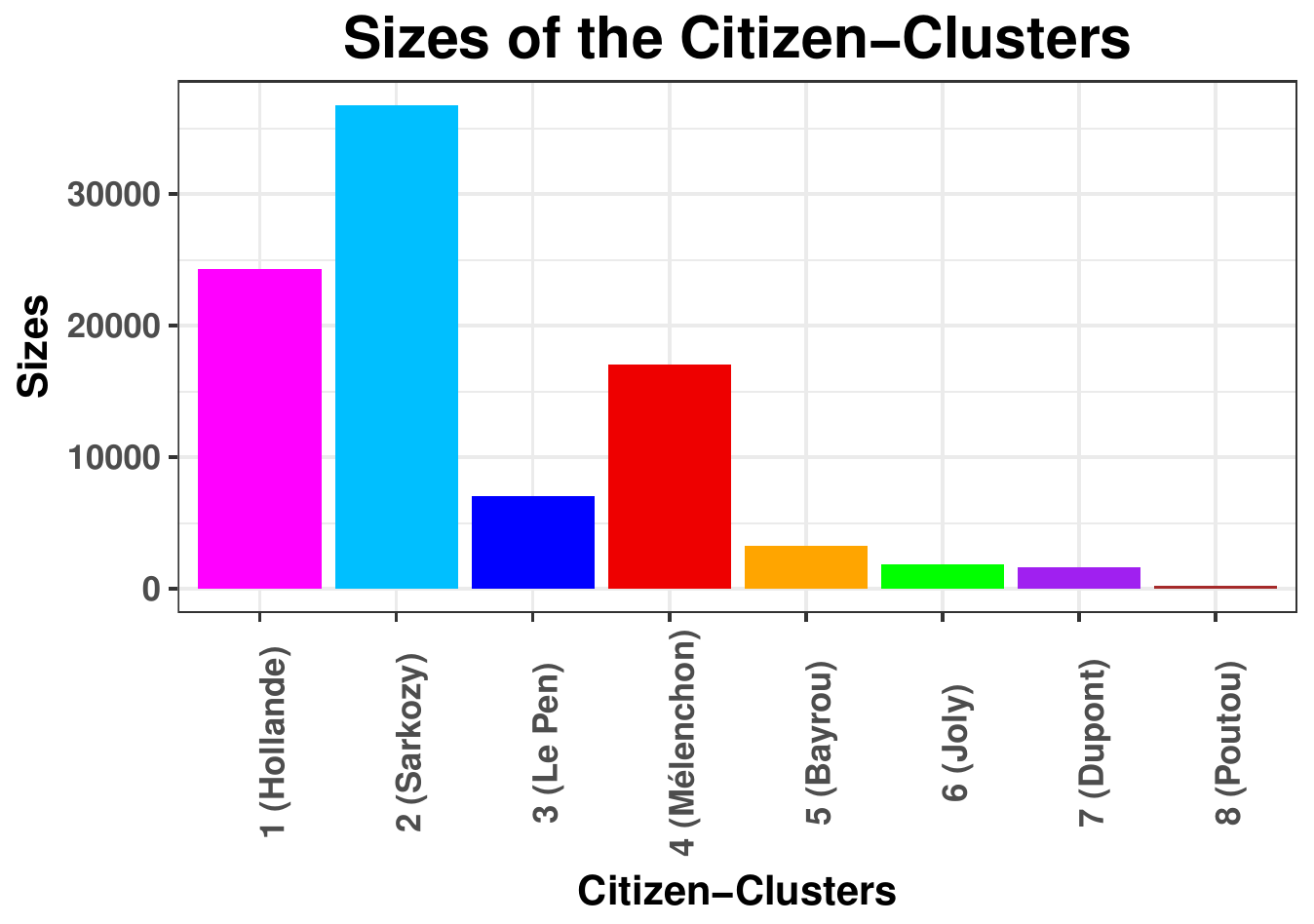}
}
\caption{\textbf{Citizen-Clusters} Figure (a) shows interactions between the citizen-clusters and candidate-walls.  The sizes of the balloons are the elements of $\Psi_C$ (defined in \eqref{def:psi_c}).  Figure (b) shows the number of citizens in each cluster. In Figure (b), we label each citizen-cluster by the corresponding candidate .  For example, the first citizen-cluster is Hollande-centered from Figure (a), so we label it as 1 (Hollande) in Figure (b).  }
\label{fig:favratio_ball_sizes}
\end{figure}

Categorizing the citizens based upon where they focus their attention produces a partition.  For any partition of citizens, $\mathscr{P}:\{1, \dots, N_C\} \rightarrow \{1, \dots, K_C\}$ where $N_C=92,226$ is the number of citizens and $K_C$ is the number of citizen-clusters, define matrix $\Psi_C\in\mathbbm{R}^{K_C\times 8}$ such that for any $a \in \{1, \dots, K_C\}$ and $b \in \{1, \dots, 8\}$,
\begin{equation}\label{def:psi_c}
[\Psi_C]_{a,b} = 
\frac{\text{$\#$ of comments from citizens in cluster $a$ under posts on $b$th candidate-wall}}{(\text{$\#$ of citizens in cluster $a$})\times (\text{$\#$ of posts on $b$th candidate-wall})}.
\end{equation}
Figure \ref{fig:favratio_ball_sizes} gives a balloon plot of $\Psi_C$ for the partition created by where citizens focus their attention.  It also shows a clear candidate-centered structure: Each candidate has a corresponding citizen-cluster that mainly comment on their posts.  Combined with the size of each citizen-cluster, it shows leading candidates attract larger clusters of citizens. See supplementary material for more evidence for candidate-centered structure.

However, such strong candidate-centered structure, where citizens primarily comment on the wall of one candidate, does not lead to the conclusion that citizens devote their attention to candidates rather than issues. It might be an ``illusion" from the ``magnifying" effect of Facebook (\cite{webster2014marketplace}).  One possibility is many citizens may only follow one candidate on Facebook, so they can only see posts from one candidate. Even if they are interested in topics that are discussed by many candidates, they are likely to comment only on the candidate's posts that they follow.  In this case, even a slight more interest in one candidate can be magnified by Facebook to a strong candidate-centered structure.  To understand whether the citizens' attention is only directed by candidates, we dig more deeply into the discussion threads in the following sections.

Importantly, the partition of citizens in Figure \ref{fig:favratio_ball_sizes}, which is created by where citizens focus their attention, uses the additional information of \textit{which of the eight candidates writes each post}. In other words, this partition of the rows of $A \in \mathbb{R}^{92,226 \times 3239}$ uses a partition of the 3239 columns of $A$ which is defined by which candidate writes the post.  The next sections will define two additional partitions of the citizens.  Neither of these partitions will use the information of which candidate writes the post. The summary $\Psi_C$ will be computed with these new partitions to help interpret whether they are discovering candidate-centered structure.

\subsection{Studying the graph using \textsc{di-sim}}\label{sec: di-sim}

\begin{figure}[H]\centering 
	\includegraphics[width=0.5\columnwidth]{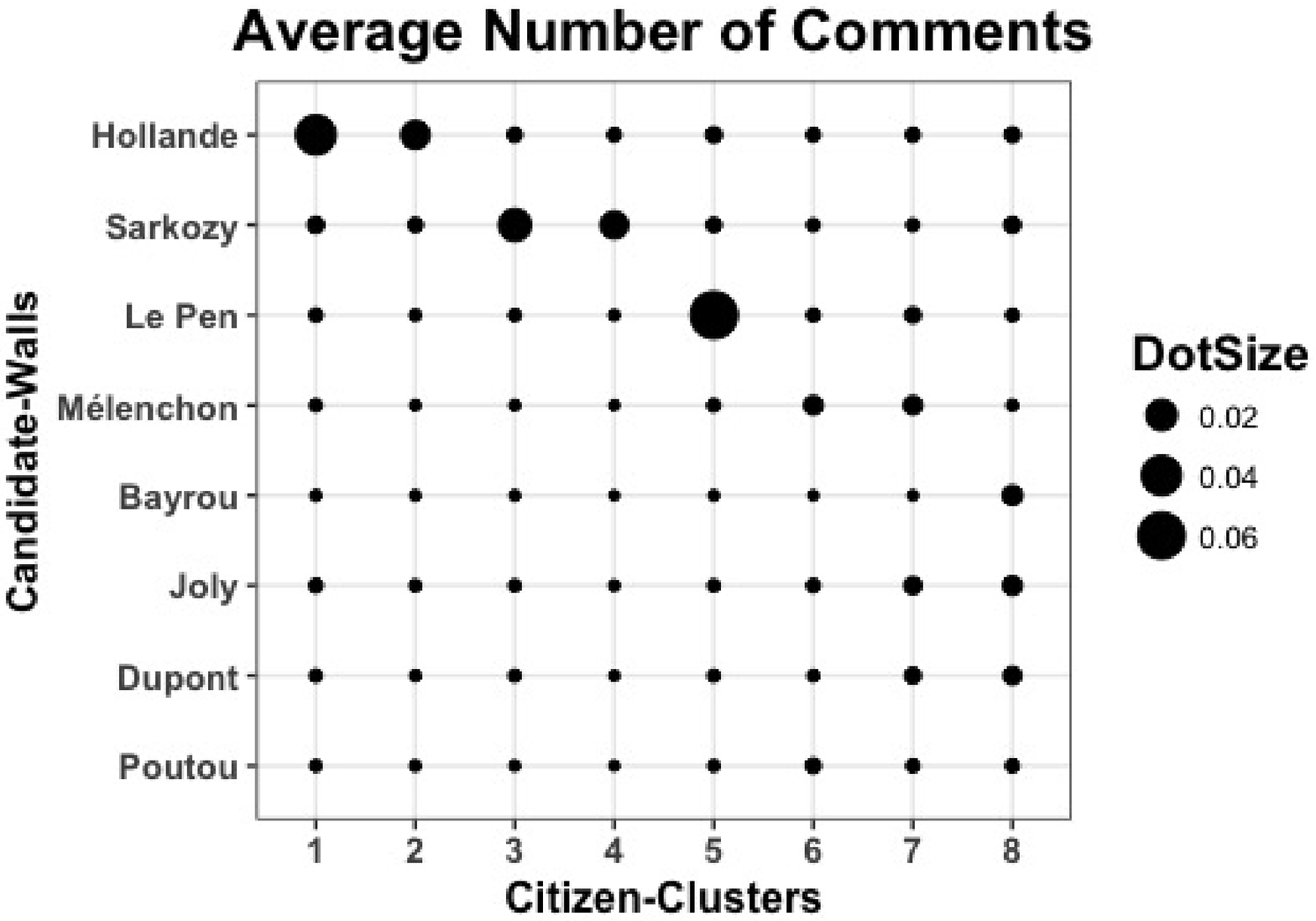}
	\caption{\textbf{The Citizen-Clusters by \emph{DI-SIM}.} Similar to Figure \ref{fig:favratio_balloon}, this figure shows the balloon plot of $\Psi_C$ corresponding to the citizen-clusters by \textsc{di-sim}.}
	\label{fig:regspec_balloon}
\end{figure}

Despite the overwhelming evidence for strong candidate-centered clusters in Figure \ref{fig:favratio_ball_sizes}, the spectral algorithm \textsc{di-sim} (\cite{rohe2016co}) finds a different partition of the citizens.  \textsc{Di-sim} partitions both citizens and candidate-posts by applying a spectral clustering algorithm.  It applies the singular value decomposition to a normalized version of the adjacency matrix $A$ (defined in \eqref{eq:adjMat})\footnote{This normalized version of the adjacency matrix $A$ is the regularized graph Laplacian which we will define in details in \eqref{def:graphLap}}. By applying k-means to the top left and right singular vectors, \textsc{di-sim} partitions the citizens and posts to different clusters.\footnote{\textsc{Di-sim} is similar to the step $\ref{svd}$ - $\ref{cluster_post}$ in Algorithm \ref{algorithm}. It applies the singular value decomposition on the graph Laplacian instead.  }  Figure \ref{fig:regspec_balloon} displays the matrix $\Psi_C$ (defined in \eqref{def:psi_c}) for the partition of citizens created by \textsc{di-sim}.  Only the top three candidates have clusters that focus on them: Hollande and Sarkozy each has two clusters and Le Pen has one cluster that focuses on her.  Other citizen-clusters (6,7,8) spread across multiple candidates.

One possible reason for the discrepancy between the attention-based partition and the partition from \textsc{di-sim} is that there may be some additional structure and \textsc{di-sim} is finding a mixture of the candidate-centered structure with that additional structure.  
\texttt{pairGraphText}, which we will introduce in the following sections, confirms that there is also an issue-centered structure in the network by incorporating text information.


\section{Graph Contextualization with \texttt{pairGraphText}}\label{spectral_contextualization}

As shown in Section \ref{background}, there are at least two good clusterings of the nodes (by attention-ratio or by \textsc{di-sim}).  Given the potentially large number of plausible clusterings of the nodes, the overarching aim of graph contextualization is to find a co-clustering of $A$ (i.e. clustering both its rows and columns) such that these clusters align with a partition in the contextualizing information.  

To quantify and utilize the contextualizing information, Section \ref{subsec:preprocess_text} describes how we preprocess the text in the discussion threads.  Section \ref{bow_mat} defines the document-term matrices to represent the text used by citizens and candidate-posts.  Section \ref{subsec:sc} introduces the \texttt{pairGraphText} algorithm.

\subsection{Preprocessing the text}\label{subsec:preprocess_text}
To preprocess the text, we represent the text in document-terms, remove numbers, symbols (e.g. \%, $@$, etc), and stop words (e.g. $\texttt{le}$, $\texttt{la}$, $\texttt{en}$, $\texttt{au}$, etc.) and transfer words into their roots by stemming.  For example, $\texttt{maintenaient}$, $\texttt{maintenait}$, $\texttt{maintenant}$, $\texttt{maintenir}$ are transferred into their root $\texttt{mainten}$.

\subsection{Document-term matrices (node covariate matrices)}\label{bow_mat}
From the cleaned text, we retain two different sets of words: ``citizen-words'' which are contained by at least $0.1\%$ of the comments, and ``thread-words'' which are contained in at least $0.1\%$ of the contents in threads (i.e. posts and comments).  In this data, over 99\% of citizen-words and thread-words are overlapped, such as \texttt{franc}, \texttt{vot}, \texttt{plus}, etc.  There are also thread-words that are not in citizen-words, such as \texttt{confrontaient}, \texttt{relancait}, etc.

To contextualize the citizens with the words that they write, define $X\in\mathbb{R}^{N_C\times M_C}$, where $N_C$ is the number of citizens and $M_C = 2020$ is the number of citizen-words.  For citizen $i$ and citizen-word $j$, $$X_{ij} = \# \text{ of comments from citizen $i$ that contain citizen-word $j$}.$$ 
Representing the candidate-posts is not as simple.  Candidate-posts provide platforms for conversations, but usually it is the comments underneath it that generate conversations. This phenomenon is colloquially referred to as ``thread highjacking,'' where the discussion thread (beneath a candidate-post) is used to discuss something other than what is discussed in the candidate-post.  In particular, many of the candidate-posts direct their followers to interviews that happen in traditional media. 
Thus, to properly contextualize the thread, one must include the text that citizens are responding to, which is not necessarily the candidate-post.  To represent the text that citizens are responding to when they post a comment in a thread, we use matrix $Y \in \mathbb{R}^{N_P\times M_P}$, where $N_P = 3239$ is the number of candidate-posts and $M_P = 2021$ is the number of thread-words.  For candidate-post $i$ and thread-word $j$, 
\begin{align*}
Y_{ij} = &\bm{1}\{\text{candidate-post $i$ contains thread-word $j$} \} + \\ &\# \text{ of comments underneath candidate-post $i$ that contain thread-word $j$}.
\end{align*}

We refer to $X$ and $Y$ document-term matrices and consider them as node covariate matrices that contain the text information about both types of nodes (citizens and candidate-posts).  The rows index the nodes (citizens or candidate-posts) and columns index the dictionaries (citizen-words or thread-words).  Our setting allows citizen-covariates and post-covariates to differ in both type and number.  In general, there could be various types of covariates.  Note that categorical covariates should be re-expressed with dummy variables.  In practice, node covariate matrices $X$ and $Y$ should be centered and scaled by column before analysis. 

\subsection{\texttt{pairGraphText}}\label{subsec:sc}

\texttt{pairGraphText} is a refinement of Covariate Assisted Spectral Clustering (CASC) \citep{binkiewicz2017covariate}.  In CASC, the graph is uni-partite.  Denote $X \in \mathbb{R}^{N \times M}$ as the node covariate matrix and $L \in \mathbb{R}^{N \times N}$ as the regularized graph Laplacian  \begin{equation}\label{def:graphLap}
L = D_C^{-1/2}AD_P^{-1/2},
\end{equation}
where $D_C$ and $D_P$ are diagonal matrices with $[D_C]_{ii} = \sum\limits_{j}A_{ij}+\tau_c$ and $[D_P]_{jj} = \sum\limits_{i}A_{ij}+\tau_p$, where $\tau_c$($\tau_p$) is set to be the average row (column) degree.  When the uni-partite graph is undirected, $D_C = D_P$.  CASC adds the covariate assisted part $C = XX^T$ to the regularized graph Laplacian and performs spectral clustering on the following similarity matrix
$$S_{casc}(h) = L+hC.$$  

To generalize CASC, \texttt{pairGraphText} refines the matrix $C$ in several ways.
This refinement will first be expressed in terms of a uni-partite graph where $X = Y$.  Replace $C=XX^T$ with $$C_W = XWX^T$$ for some matrix $W$.  Note that when $W$ is identity matrix, $C_W = C$.  By imposing matrix $W$, \texttt{pairGraphText} addresses the following limitations of CASC.  

\begin{itemize}
	\item   For any matrix $H$, denote its $i$th row as $H_{i\cdotp}$ and its $j$th column as $H_{\cdotp j}$.  Note that $C_W = \sum_{ij}W_{ij}X_{\cdotp i}X_{\cdotp j}^T$.  So, when $W_{ij}$ is nonzero for $i\not= j$, it creates an ``interaction" between $X_{\cdotp i}$ and $X_{\cdotp j}$, i.e. $i$th and $j$th covariates.  Such interactions are not included in  $C = XX^T = \sum_{j}X_{\cdotp j}X_{\cdotp j}^T$.
	\item In $C$, there is not a natural way of excluding covariates, i.e. discarding columns of $X$.  However, in many settings, several covariates could be unaligned with the graph and they should be excluded from the similarity matrix.  $C_W$ can select covariates by setting some elements (or rows/columns) of $W$ to zero.  
	\item $C$ presumes that two nodes are more likely to be connected when they have similar covariates.  But in some situations, this is not true.  For example, in a dating network, relationships are more prevalent among men and women than two people of the same gender.  In $C_W$, if $W_{ii}$ is negative, then two nodes are closer in the similarity matrix $C_W$ if they have different values for the $i$th covariate.  
	\item  The symmetric matrix $C$ only allows for symmetric contributions of covariates, which may not be the case for directed graphs. This can be addressed by allowing $W$ to be asymmetric.  
    \item Finally, CASC was not designed for bi-partite networks.  In a bipartite graph, the rows of $A$ might have different contextualizing measurements than the columns of $A$.  In the Facebook data, these measurements correspond to the matrices $X$ and $Y$.  Because they have a different number of measurements, the multiplication $XY^T$ is not defined for the Facebook data.  However, the multiplication $XWY^T$ is well defined for a rectangular $W$.  This removes the need for a one-to-one correspondence between the columns of $X$ and $Y$; they could contain entirely different types of measurements.  
\end{itemize}

We propose estimating a matrix $W$ to address the issues above.  
Define the \textbf{call-response matrix} 
\begin{equation}\label{def:w}
W=X^TLY,
\end{equation} 
which measures the correlation between thread-words and citizen-words \textit{along the graph}.  For example, if discussion threads containing the word \texttt{franc} have comments from citizens that are likely to say \texttt{vot}, then citizen-word \texttt{vot} is highly correlated with a thread-word \texttt{franc} along the graph. 

\begin{figure}[H]\centering 
\subfigure[][]{\label{fig:w_part}
\includegraphics[width=0.49\columnwidth]{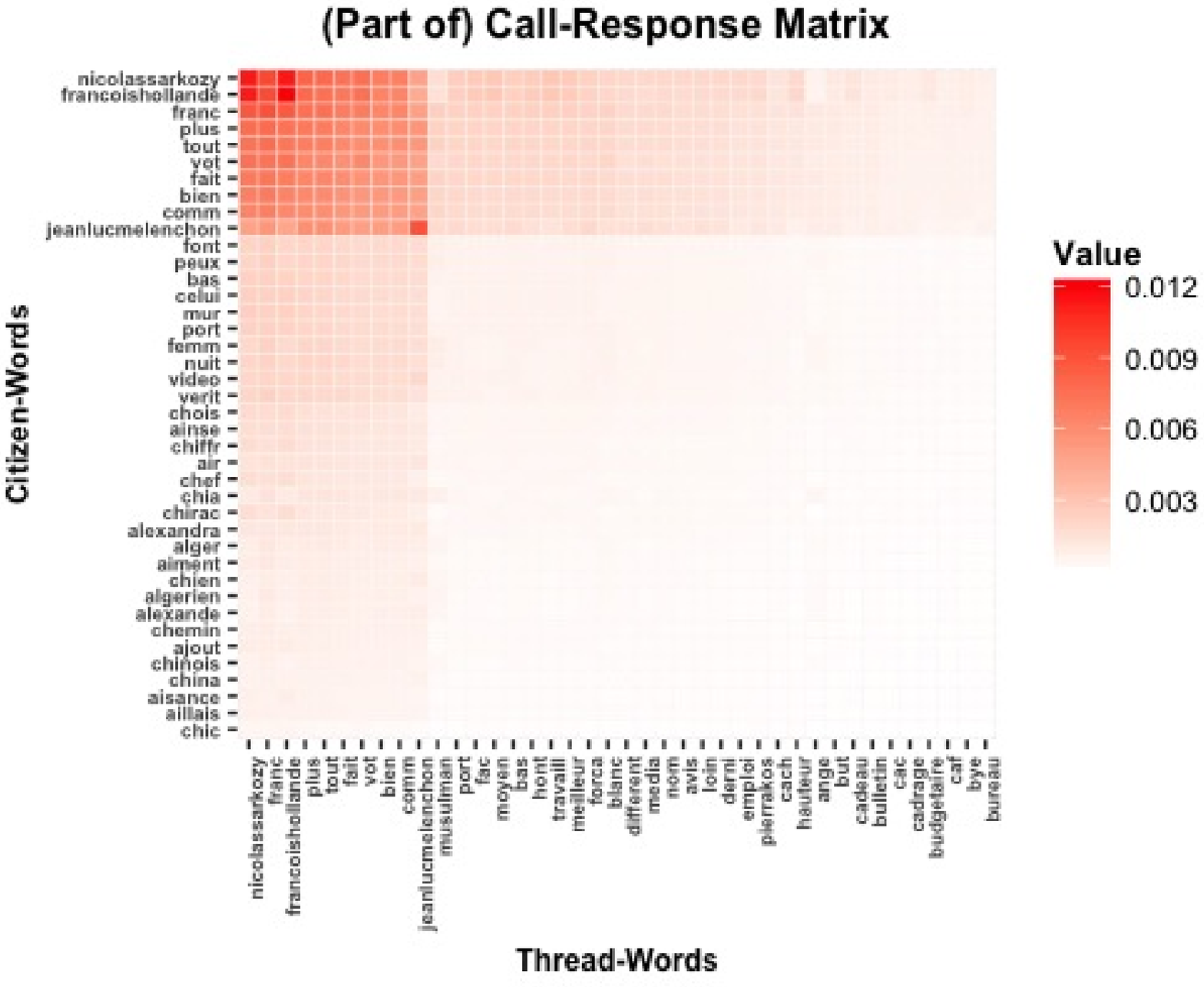}}
\subfigure[][]{\label{fig:w_part_thresh}
\includegraphics[width=0.49\columnwidth]{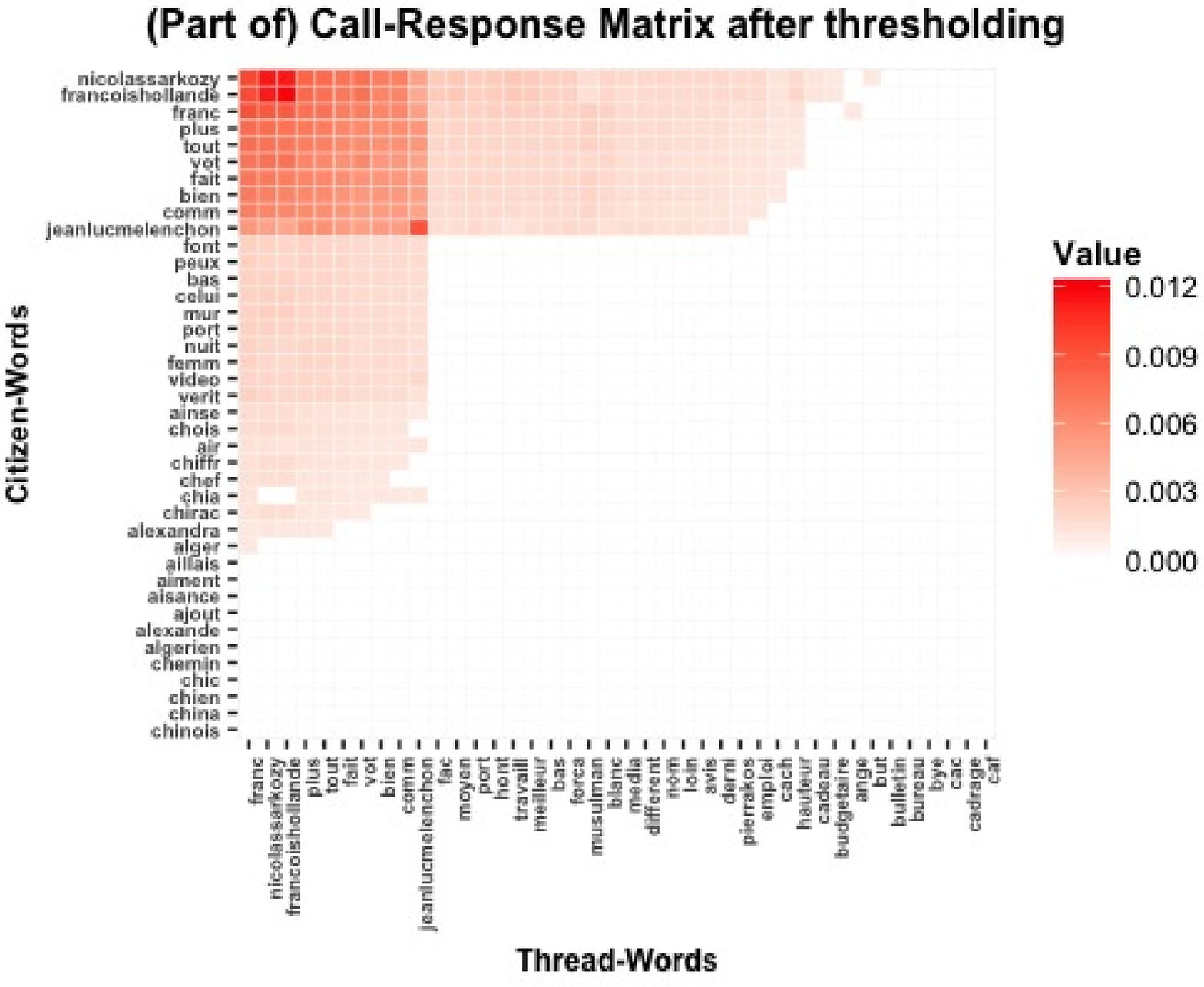}}
\caption{\textbf{Part of the Call-Response Matrix before and after Thresholding}  Some pairs of words are relatively more highly correlated, like $\texttt{nicolassarkozy}$ and $\texttt{francoishollande}$, $\texttt{jeanlucmelenchon}$ and $\texttt{jeanlucmelenchon}$, $\texttt{vot}$ and $\texttt{franc}$, etc. After thresholding, only the relatively highly correlated pairs of words are left, making the call-response matrix much more sparse.}
\label{fig:w}
\end{figure}

To illustrate $W=X^TLY$, examine a single element $x^T L y$, where $x \in \mathbb{R}^{92,226}$ is a column of $X$ corresponding to word \texttt{vot} and $y \in  \mathbb{R}^{3239}$ is a column of $Y$ corresponding to word \texttt{franc}.  So,  $x_i$ is the number of times that citizen $i$ uses \texttt{vot} and $y_j$ is the number of times that \texttt{franc} appears in the thread for candidate-post $j$.  If $x$ is centered and independent of $L$ and $y$, then $x$ is an uninformative covariate, and $\mathbbm{E}[x^T L y] = \mathbbm{E}(\mathbbm{E}(x^T|L,y) L y)  = 0$.  Conversely, if for centered $x$ and $y$,
\[x^TLy = \sum_{i,j: A_{ij} = 1} \frac{x_i y_j}{\sqrt{[D_C]_{ii}[D_P]_{jj}}}\]
is large (positive or negative), it suggests that linked nodes in $L$ have (positively or negatively) correlated values of $x$ and $y$.  Figure \ref{fig:w} gives a small part of the call-response matrix.

There are thousands of words in the discussion threads.  To select the highly correlated words along the graph, we define a hard-threshold function on $W$, 
\begin{equation}\label{eq:thresh}
[T_{\omega}(W)]_{sr} = 
\begin{cases} 
W_{sr}, & \text{if $W_{sr}>\omega$}\\
0, & \text{o.w.}
\end{cases}
\end{equation}
In practice, we can set the threshold $\omega$ as the $1-\alpha$ quantile of $|W_{ij}|$'s.

Thus, we finally define the matrix that replaces $C$ from CASC.  For \texttt{pairGraphText}, define
\begin{equation}\label{def:text_assisted}
C_T = XT_{\omega}(W)Y^T.
\end{equation}

The following diagram reviews how \texttt{pairGraphText} refines the matrix $C$ from CASC.  
\begin{center}
\begin{tikzpicture}[
 every matrix/.style={ampersand replacement=\&,column sep=3cm,row sep=0.2cm},
c/.style={draw,thick,rounded corners,fill=yellow!20,inner sep=.1cm},
to/.style={->,>=stealth',shorten >=1pt,thick},
every node/.style={align=center}]

  \matrix{
      \node[c] (c1) { $XX^T$
       };
      \& \node[c] (c2) {$XWX^T$
      }; 
      \& \node[c] (c3) {$XWY^T$
   }; 
      \&  \node[c] (c4) {$XT_{\omega}(W)Y^T$
       };  \\
  };

   \draw[to] (c1) -- node[midway,above] {create\\interactions}
  node[midway,below] {with $W$} (c2);
   \draw[to] (c2) -- node[midway,above] {allow for different\\ citizen- and post-}
node[midway,below] {covariates} (c3);
   \draw[to] (c3) -- node[midway,above] {select interactions\\ by thresholding}
node[midway,below] {elements of $W$} (c4);

 \end{tikzpicture}
\end{center}

Note that 
\[C_T  = \sum\limits_{ij}[T_{\omega}(W)]_{ij}X_{\cdotp i}Y_{\cdotp j}^T\] 
shows closeness of citizens and candidate-posts based on their usage of words in the network.  $[C_T]_{ij}$ is large when citizen $i$ and candidate-post $j$ use many highly correlated pairs of words.  The threshold function $T_{\omega}(\cdotp)$ helps select pairs of words, and imposes sparsity when $W$ is high-dimensional.    

Therefore, \texttt{pairGraphText} applies \textsc{di-sim} to the similarity matrix:
\begin{equation}\label{def:L}
S = L+hC_T.
\end{equation}
This similarity matrix combines both the graph information, represented by $L$, and the text information, represented by $C_T = XT_{\omega}(W)Y^T$, with a tuning parameter $h$ to balance between these two parts.   

\begin{algorithm}
	\caption{\texttt{pairGraphText}}
	\label{algorithm}
	Input: adjacency matrix $A\in \mathbbm{R}^{N_P\times N_C}$, node covariate matrices $X\in \mathbbm{R}^{N_P\times M_P}$ and $Y\in\mathbbm{R}^{N_C\times M_C}$, number of citizen-clusters $K_C$, number of post-clusters $K_P$, weight $h$, and the significance level $\alpha$.  
	\begin{enumerate}
		\item Compute the regularized graph Laplacian $L$ from $A$ as in \eqref{def:graphLap}. Center $X$ and $Y$ by column. (In practice, scaling $X$ and $Y$ by rows and columns or using weighted $X$ and $Y$ might also be beneficial.  See more details in Section \ref{sec: compare_different_scale}. )
		\item Compute $W = X^TLY$. Set $\omega$ to be the $1-\alpha$ quantile of $|W_{ij}|$'s.
		\item Compute the similarity matrix for \texttt{pairGraphText} as $$S = L+hXT_{\omega}(W)Y^T.$$
		\item\label{svd} Compute the top $K$ left and right singular vectors $U_C\in\mathbb{R}^{N_C\times K}$, $U_P\in\mathbb{R}^{N_P\times K}$ corresponding to the $K$ largest singular values of $S$, where $K=\min\{K_C,K_P\}$.  
		\item Form matrices $U_C^{\ast}\in\mathbb{R}^{N_C\times K}$ and $U_P^{\ast}\in\mathbb{R}^{N_P\times K}$ such that for any $i\in\{1,\dots, N_C(N_P)\}$,
		\begin{equation}\label{def:ucup}
		[U_C^{\ast}]_{i\cdotp} = \frac{[U_C]_{i\cdotp}}{\|[U_C]_{i\cdotp}\|_2}
		\text{ and }
		[U_P^{\ast}]_{i\cdotp} = \frac{[U_P]_{i\cdotp}}{\|[U_P]_{i\cdotp}\|_2}.
		\end{equation}
		\item\label{cluster_citizen} Cluster the rows of $U_{C}^{\ast}$ into $K_C$ clusters with k-means. If the $i$th row of $U_{C}^{\ast}$ falls in the $k$th cluster, assign citizen $i$ to citizen-cluster $k$.  
		\item\label{cluster_post} Cluster the candidate-posts by performing step $\ref{cluster_citizen}$ on the matrix $U_{P}^{\ast}$ with $K_P$ clusters.  
	\end{enumerate}
\end{algorithm}

\section{Issue-centered structure}\label{issue_structure}

We identify topics that attract public's attention in the Facebook discussion threads using \texttt{pairGraphText}.  We scale the document-term matrices by both rows and columns.\footnote{We replace $X_{ij}$ and $Y_{ij}$ by $X_{ij}/\sqrt{\sum\limits_i X_{ij} \sum\limits_j X_{ij}}$ and $Y_{ij}/\sqrt{\sum_i Y_{ij}\sum_j Y_{ij}}$.}  From the scree plot of the singular values of $S$ (see Figure 3 in supplementary material), we decide to study the top $K=4$ clusters due to the large gap after the fourth singular value.  To study how the text in discussion threads affects the partition of citizens and candidate-posts, we show the clustering results in three cases: (i) when we use no text, i.e. the tuning parameter $h$ in Equation \eqref{def:L} is $h = 0$,\footnote{When $h = 0$, \texttt{pairGraphText} is equivalent to \textsc{di-sim}.} (ii) when we incorporate text, i.e. $h = 0.035$,\footnote{In case (ii), $h$ can be any real positive value.  We choose $h = 0.035$ since it shows clusters with major differences from both cases when $h = 0$ and when $h = \infty$.  Recall the similarity matrix $S = L+hC_T$ (see \eqref{def:L}).  For identification of $h = 0.035$, we scale the text-assisted part $C_T$ to have the same second singular value with $L$. Then, $h$ means how much we weigh the text-assisted part in \texttt{pairGraphText}. $h = 0.035$ means that we weigh the text-assisted part 0.035 times of the graph information.}   and (iii) when we only use the text assisted part (defined in $\eqref{def:text_assisted}$), i.e. $h = \infty$.  

Section \ref{sec: clusters_diff_h} shows that with more text incorporated (i.e. with larger $h$), the clusters become less candidate-centered.  Section \ref{sec: word_content} introduces a word-content strategy to extract topics of clusters.  Section \ref{sec: topics} describes the cluster topics and supports Section \ref{sec: clusters_diff_h} by showing that clusters with larger $h$ are more heavily focused on the contextualizing information.

\subsection{The clusters from \texttt{pairGraphText} with larger $h$ are less candidate-centered}\label{sec: clusters_diff_h}
\begin{figure}[H]
\subfigure[][]{\label{fig:postclus_candwall_balloon}
\includegraphics[width=0.7\columnwidth]{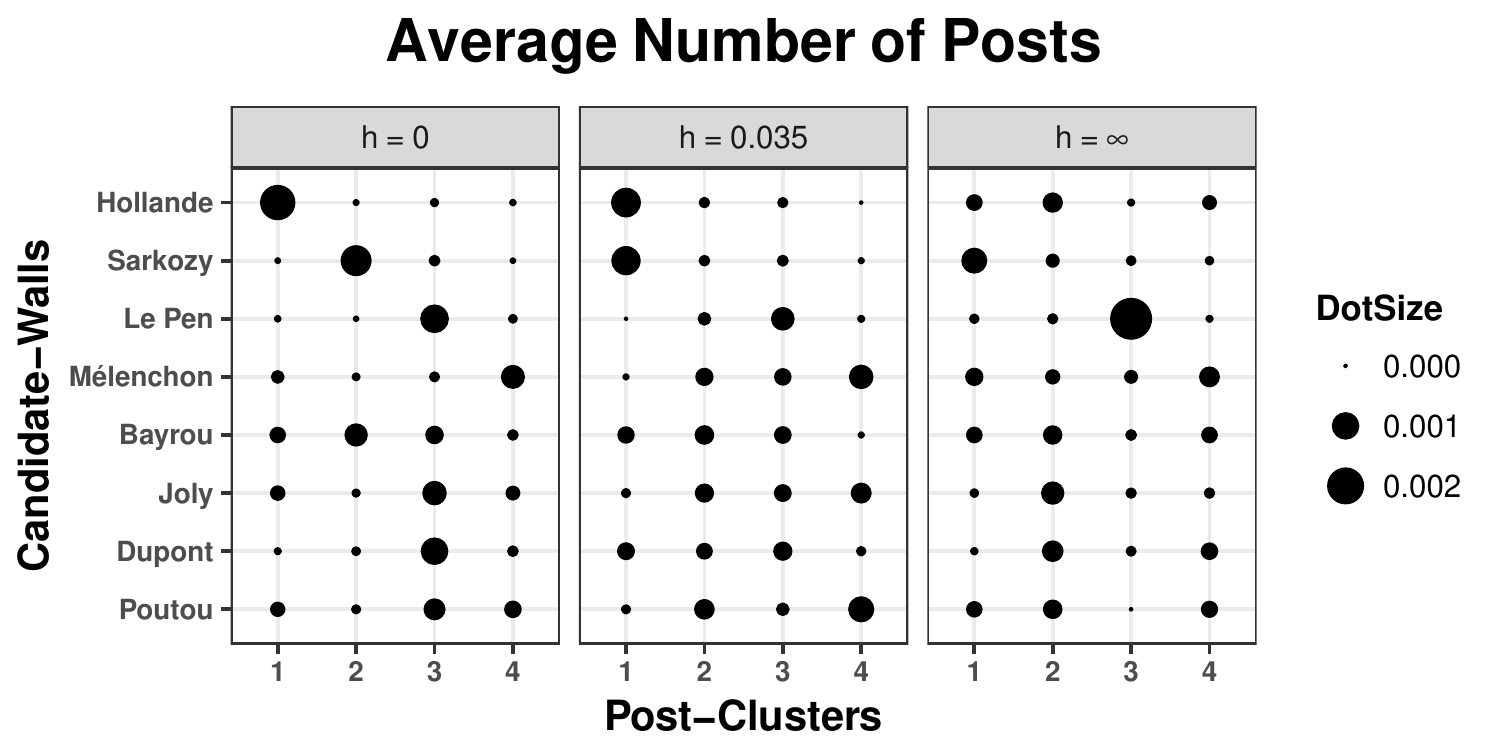}
}
\subfigure[][]{\label{fig:citizenclus_candwall_balloon}
\includegraphics[width=0.7\columnwidth]{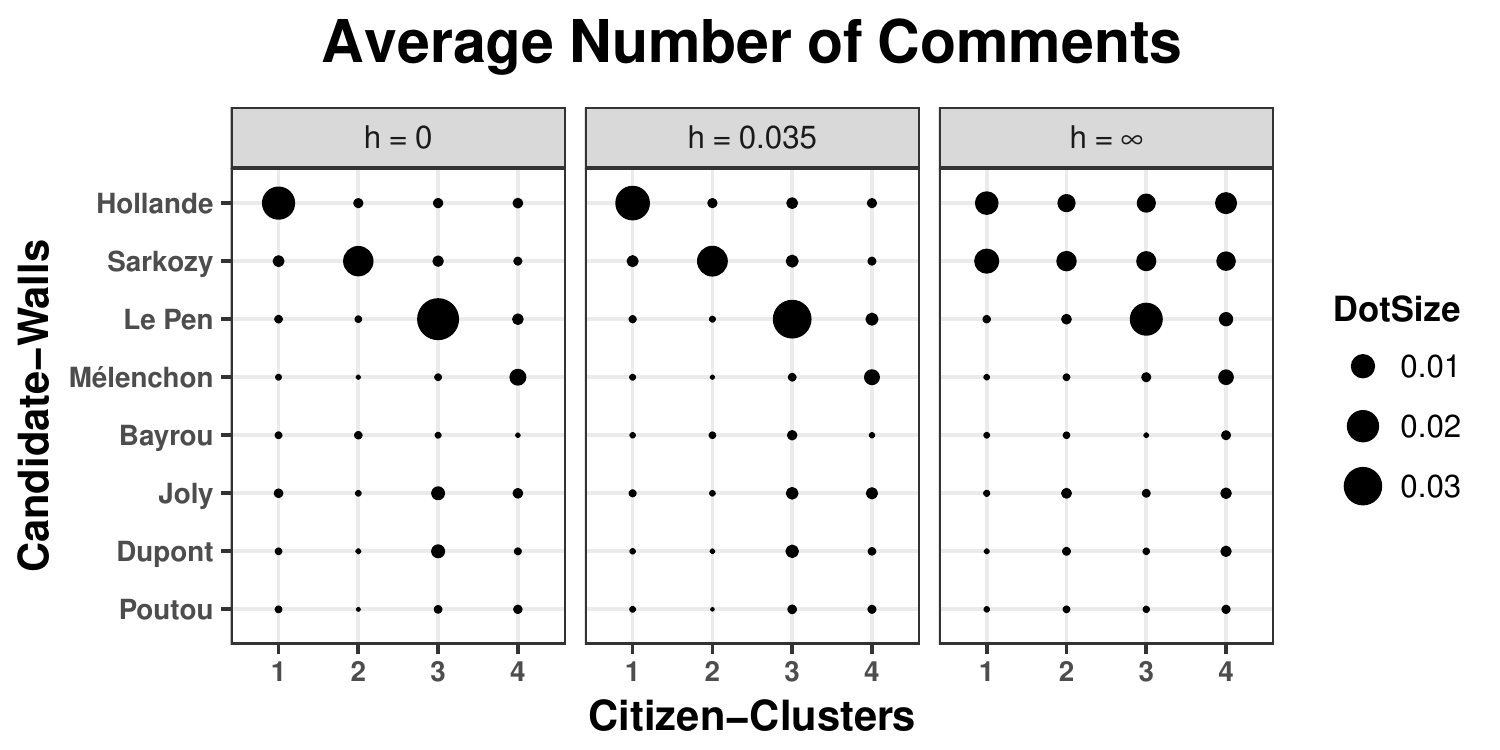}
}
\caption{\textbf{Clusters and Candidate-Walls} Figure (a) and (b) display $\Psi_P$ and $\Psi_C$ in ballloon plots for the three cases.}
\label{fig:clus_candwall_balloon}
\end{figure}

For each partition of candidate-posts, $\mathcal{P}:\{1,...,N_P\}\rightarrow\{1,...,4\} $, we define the matrix $\Psi_P\in \mathbbm{R}^{4\times 8}$ such that for any $a\in\{1,...,4\}$ and $b\in\{1,...,8\}$,
\begin{equation}\label{def:psi_p}
[\Psi_P]_{ab} = \frac{\text{\# of posts in cluster $a$ from candidate $b$'s wall}}{(\text{\# of posts in cluster $a$}) \times (\text{\# of posts from candidate $b$'s wall})}.
\end{equation}
$\Psi_P$ shows how post-clusters distribute on candidate-walls.  This is similar to $\Psi_C$ defined in $\eqref{def:psi_c}$, which shows how citizen-clusters interact with candidate-walls.  Figure \ref{fig:clus_candwall_balloon} displays $\Psi_P$ and $\Psi_C$ in balloon plots in the three cases. When we use no text, i.e. $h = 0$, there appears some candidate-centered structure in both citizen-clusters and post-clusters. As we incorporate text, in the case when $h = 0.035$, each post-cluster spreads across multiple candidates. With even more text incorporated, in the case $h=\infty$, neither of the post-clusters nor citizen-clusters are candidate-centered.  In the following subsections, we identify the cluster topics using key words, comments and posts.

\subsection{A word-content strategy to identify cluster topics}\label{sec: word_content}
To identify the cluster topics, we first identify \texttt{keywords} for each cluster, which we will define in Section \ref{sec: sigwords}.  These keywords give the first impression of the cluster topics.  

However, it is insufficient to examine the words in isolation, because the same word is often used differently by different subsets of the population.  For example, \texttt{religion} is often used by citizens talking about the \texttt{religion of peace} and it is also often used by atheists criticizing its appearance in the public sphere.  Thus, to identify the cluster topics, besides identifying keywords, we also need to read through the conversations that contain these keywords.  We focus on the \texttt{central conversations} in each cluster, which we will define in Section \ref{sec: centrality}.  

We call this strategy \texttt{word-content strategy}, where for each cluster, we (i) identify the keywords and (ii) read through the central conversations that contain the keywords in the cluster. 


\subsubsection{Identify the keywords}\label{sec: sigwords}
We identify the keywords in each cluster by setting ``scores".  For any $k\in\{1,\dots,4\}$ and $j\in \{1,\dots, M_C\}$, define the \texttt{score} of citizen-word $j$ in citizen-cluster $k$ as
\begin{align*}
\Phi_{kj}=\frac{\sum\limits_{i\in k}X_{ij}}{\sum\limits_{i\in k}\hat{X}_{ij}},
\text{ where } \hat{X}_{ij} = \frac{\sum\limits_{j}X_{ij}\sum\limits_{i}X_{ij}}{\sum X_{ij}},
\end{align*}
and $i\in k$ denotes the citizen $i$ belongs to cluster $k$.  We similarly define the scores of thread-words in post-clusters based on the document-term matrix of candidate-posts $Y$.  These scores are also discussed in \cite{witten2011classification}, where they are derived by maximum likelihood on a Poisson model.  We define the \texttt{keywords} in a cluster to be the words with the largest scores in the cluster.  We show keywords of each cluster in Section \ref{sec: topics}.

\subsubsection{Identifying central conversations}\label{sec: centrality}
We identify the central conversations by diagnostics from k-means clustering.  Recall the \texttt{pairGraphText} algorithm partitions citizens by applying k-means on the $N_C$ rows of matrix $U_C^{\ast}\in\mathbbm{R}^{N_C\times 4}$ (defined in \eqref{def:ucup}) which correspond to the $N_C$ citizens.  For any citizen $i$, we denote their \texttt{cluster-centrality} as $$\rho_i = [U_C^{\ast}]_{i\cdotp}^T[\mu_C^{\ast}]_i,$$
where $[\mu_C^{\ast}]_i$ is the cluster centroid of citizen $i$ from k-means on rows of $U_C^{\ast}$.  There are four different cluster centroids.  For each cluster, the \texttt{central citizens} are the citizens in the cluster with the largest cluster-centrality, i.e. those that align best with the cluster centroid.  We similarly define the \texttt{central posts} for post-clusters.  For a citizen-cluster, the \texttt{central conversations} are the comments from the central citizens;  for a post-cluster, the \texttt{central conversations} are the discussion threads (including posts and comments) initiated by the central posts.

We read through the central conversations that contain the keywords in each cluster.  This word-content strategy helps us identify topics that attract citizens' attention.  We will show these topics in Section \ref{sec: topics}.

\subsection{Topics of clusters}\label{sec: topics}
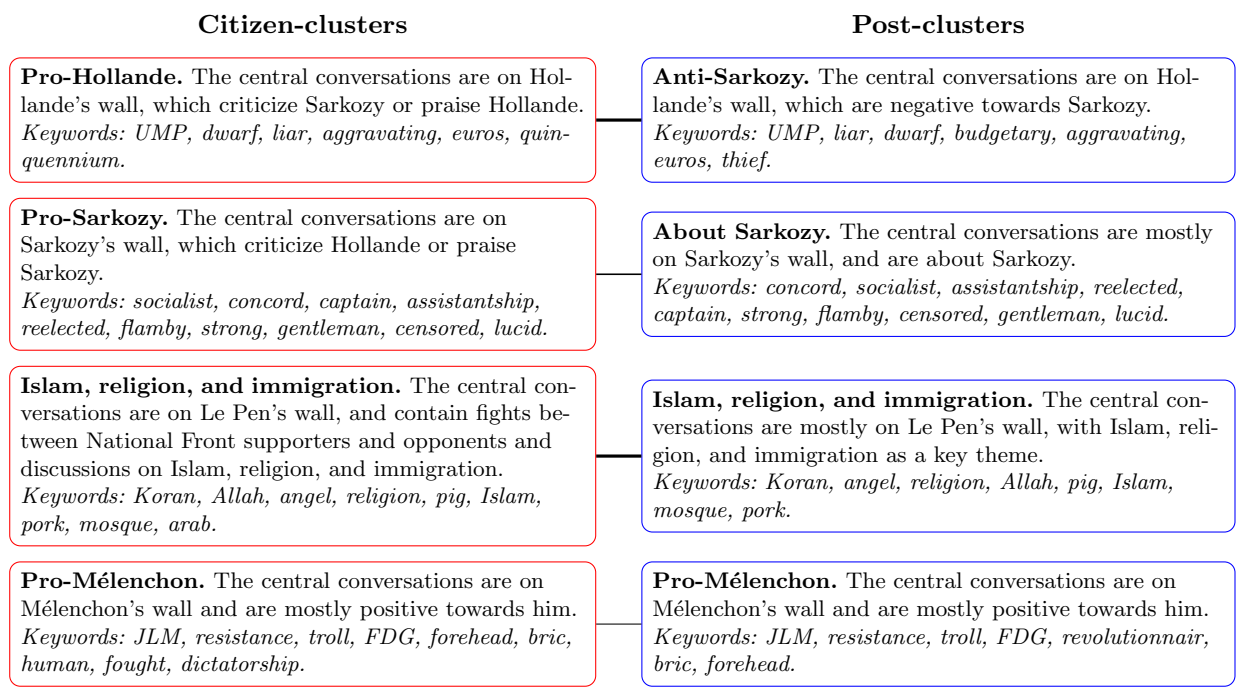
\begin{figure}[H]
\begin{center}
\begin{tikzpicture}[
every matrix/.style={ampersand replacement=\&,column sep=0.3cm,row sep=0.2cm},
citizen/.style={draw=red,thin,rounded corners,
inner sep=.15cm,text width = 7.9cm, scale = 0.95},
post/.style={draw=blue,thin,rounded corners,
inner sep=.15cm,text width = 8cm, scale = 0.95}]
  \matrix{
  	   \node[scale = 1.1](t1){\textbf{Citizen-clusters}}; \&;
  	   \& \node[scale = 1.1](t2){\textbf{Post-clusters}};\\
       \node[citizen] (c1)
{\textbf{Pro-Hollande.}  The central conversations are on Hollande's wall, which criticize Sarkozy or praise Hollande. \\
\textit{Keywords: 
UMP, dwarf, liar, aggravating, euros, quinquennium.}
       };\&;
       
      \& \node[post] (p1) {\textbf{Anti-Sarkozy.}  The central conversations are on Hollande's wall, which are negative towards Sarkozy.\\   
\textit{Keywords: UMP, liar, dwarf, budgetary, aggravating, euros, thief.}
      }; \\

       \node[citizen] (c2)
{\textbf{Pro-Sarkozy.}  The central conversations are on Sarkozy's wall, which criticize Hollande or praise Sarkozy. \\ 
\textit{Keywords: socialist, concord, captain, assistantship, reelected, flamby, strong, gentleman, censored, lucid.}
   }; \&;
       \& \node[post] (p2) {\textbf{About Sarkozy.}   The central conversations are mostly on Sarkozy's wall, and are about Sarkozy. \\
\textit{Keywords: concord, socialist, assistantship, reelected, captain, strong, flamby, censored, gentleman, lucid.}
       }; \\
   
    \node[citizen] (c3) {\textbf{Islam, religion, and immigration.}  The central conversations are on Le Pen's wall, and contain fights between National Front supporters and opponents and discussions on Islam, religion, and immigration.\\      
\textit{Keywords: Koran, Allah, angel, religion, pig, Islam, pork, mosque, arab.}
   }; \&;
   \& \node[post] (p3) {\textbf{Islam, religion, and immigration.}  The central conversations are mostly on Le Pen's wall, with Islam, religion, and immigration as a key theme. \\
\textit{Keywords: Koran, angel, religion, Allah, pig, Islam, mosque, pork.}
   }; \\

 \node[citizen] (c4) {\textbf{Pro-M\'elenchon.}  The central conversations are on M\'elenchon's wall and are mostly positive towards him.\\ 
\textit{Keywords: 
JLM, resistance, troll, FDG, forehead, bric, human, fought, dictatorship.}
}; \&;
\& \node[post] (p4) {\textbf{Pro-M\'elenchon.}  The central conversations are on M\'elenchon's wall and are mostly positive towards him. \\ 
\textit{Keywords: 
JLM, resistance, troll, FDG, revolutionnair, bric, forehead.}
}; \\
  };

  \path[draw, line width = 1.2pt] (c1.east) -- (p1.west);
  \path[draw, line width = 0.6pt] (c2.east) -- (p2.west);
  \path[draw, line width = 1.2pt] (c3.east) -- (p3.west);
  \path[draw, line width = .4pt] (c4.east) -- (p4.west);

 \end{tikzpicture}
\end{center}
\caption{\textbf{Cluster topics when $\bm{h = 0}$}}
\label{fig:diagram1}
\end{figure}

We extract topics of the clusters by the word-content strategy in three cases, $h = 0$, $h = 0.035$, and $h = \infty$.  Figure \ref{fig:diagram1}, \ref{fig:diagram2} and \ref{fig:diagram3} show the cluster topics with the keywords and a brief description of the central conversations in each cluster.  In these figures, the links indicate major interactions\footnote{We only display the links that correspond to the three or four largest elements of $\Psi$ in each case.} between citizen-clusters and post-clusters, with the link widths proportional to elements of matrix $\Psi\in\mathbbm{R}^{4\times 4}$, where for any $a,b\in\{1,\dots,4\}$,
$$
\Psi_{ab} = \frac{\text{\# of comments from citizens in citizen-cluster $a$ under candidate-posts from post-cluster $b$}}{(\text{\# of citizens in citizen-cluster $a$}) \times (\text{\# of candidate-posts in post-cluster $b$})}.
$$
This is similar to matrices $\Psi_C$ defined in $\eqref{def:psi_c}$ and $\Psi_P$ defined in $\eqref{def:psi_p}$, which show how clusters (for citizens or candidate-posts) distribute on the eight candidate-walls. $\Psi$ shows how the citizen-clusters interact with the post-clusters.

When $h = 0$ (see Figure \ref{fig:diagram1}), clusters focus on candidates or the radical discussions. As we incorporate the text, in the case when $h = 0.035$ (see Figure \ref{fig:diagram2}), the citizen-clusters are similar to those when $h = 0$, but there appears a post-cluster about ecology. As we incorporate more text, in the case when $h = \infty$ (see Figure \ref{fig:diagram3}), we identify more topics, such as economic and crises.  There also appear a cluster for both citizens and candidate-posts with many copy-paste comments.  More data analysis results are in a Shiny App available at \url{https://yilinzhang.shinyapps.io/FrenchElection}.

Incorporating the text makes the central conversations more vivid representations of the clusters, allowing for a more precise interpretation of the topic.  During the 2012 French election, the citizens devoted their attention and expression in (i) the debates and fights among different candidates, (ii) radical discussions on Islam, religion, and immigration, and (iii) other topics including ecology, economy, and crises.  

\vspace{-0.5cm}
\begin{figure}[H]
\begin{center}
	\begin{tikzpicture}[
every matrix/.style={ampersand replacement=\&,column sep=0.3cm,row sep=0.2cm},
citizen/.style={draw=red,thin,rounded corners,
inner sep=.15cm,text width = 7.9cm, scale = 0.95},
post/.style={draw=blue,thin,rounded corners,
inner sep=.15cm,text width = 8cm, scale = 0.95}]
	
	\matrix{
  	   \node[scale = 1.1](t1){\textbf{Citizen-clusters}}; \&;
  	   \& \node[scale = 1.1](t2){\textbf{Post-clusters}};\\
		\node[citizen] (c1){\textbf{Pro-Hollande.}  The central conversations are on Hollande's wall, which criticize Sarkozy or praise Hollande. \\ 
\textit{Keywords: dwarf, liar, aggravating, euros, quinquennium, modest.} 
		};\&;
		\& \node[post] (p1){\textbf{Hollande vs Sarkozy.}  The central conversations are on Hollande's, Sarkozy's and Bayrou's walls, which focus on on-going debate and fights between pro-Sarkozy and pro-Hollande.  
        \\  
\textit{Keywords: residential, ancestry, chic, IRS, balance sheet, pent, loss making.} 
		}; \\
		\node[citizen] (c2){\textbf{Pro-Sarkozy.}  The central conversations are on Sarkozy's wall, which criticize Hollande or praise Sarkozy.\\ 
\textit{Keywords: socialist, concord, assistantship, captain, reelected, flamby, strong, gentleman, censored.} 
		}; \&;
		\& \node[post] (p2) {\textbf{Ecology.}  The central conversations are on Bayrou's, Joly's, and Dupont-Aignan's walls.  Ecology is discussed along with Joly and the Green party. \\ 
\textit{Keywords: 
ecologic, green, sincerity, madam, antisemitic, admired, supported, standing.}
}; \\
		
		\node[citizen] (c3) {\textbf{Islam, religion, and immigration.}  The central conversations are mostly on Le Pen's wall, which contain 
discussions on Islam, religion, and immigration. \\ 
\textit{Keywords: Koran, angel, pig, Allah, religion, Islam, pork, mosque, arab.}
}; \&;
		\& \node[post] (p3) {\textbf{Islam, religion, and immigration.}  The central conversations are mostly on Le Pen's wall, which contain discussions on Islam, religion, and immigration.\\ 
\textit{Keywords: Koran, angel, Allah, religion, pig, Islam, pork, mosque, arab.}
}; \\
		
		\node[citizen] (c4) {\textbf{Pro-M\'elenchon.}  The central conversations are on M\'elenchon's wall and are mostly positive towards him. \\
\textit{Keywords: JLM, troll, FDG, forehead, human, bric, fought, revolutionary, fraternity.}
}; \&;
		\& \node[post] (p4) {\textbf{Pro-M\'elenchon.}  The central conversations are on M\'elenchon's wall and are mostly positive towards him.  There is bigger focus on defending M\'elenchon than $h = 0$.\\
\textit{Keywords: JLM, resistance, troll, FDG, bric, revolutionary, forehead, fought, human, fraternity.}
}; \\
	};

	\path[draw, line width = 0.6pt] (c3.east) -- (p3.west);
	\path[draw, line width = 1.1pt] (c4.east) -- (p4.west);
	\path[draw, line width = 1.1pt] (c1.east) -- (p1.west);
	\path[draw, line width = 0.7pt] (c2.east) -- (p1.west);

	\end{tikzpicture}
\end{center}
\caption{\textbf{Cluster patterns when $\bm{h = 0.035}$}}
\label{fig:diagram2}
\end{figure}
\vspace{-1cm}
\begin{figure}[H]
\begin{center}
	\begin{tikzpicture}[
every matrix/.style={ampersand replacement=\&,column sep=0.3cm,row sep=0.2cm},
citizen/.style={draw=red,thin,rounded corners,
inner sep=.15cm,text width = 8cm, scale = 0.95},
post/.style={draw=blue,thin,rounded corners,
inner sep=.15cm,text width = 7.9cm, scale = 0.95}]

	\matrix{
  	   \node[scale = 1.1](t1){\textbf{Citizen-clusters}}; \&;
  	   \& \node[scale = 1.1](t2){\textbf{Post-clusters}};\\
		\node[citizen] (c1){\textbf{Hollande vs Sarkozy.} The central conversations are on Hollande's and Sarkozy's walls, which contain fights between pro-Hollande and pro-Sarkozy and are more offensive than $h = 0.035$. \\
\textit{Keywords: 
Fran\c cois Hollande, Nicolas Sarkozy, fail, president, live, incompetent, May, charisma, arrogant, dwarf, goodbye, liar.} 
		};\&;
		\& \node[post] (p1){\textbf{Pro-Sarkozy.} The central conversations are on Hollande's and Sarkozy's walls, which focus on criticizing Hollande or praising Sarkozy.\\
\textit{Keywords: 
concord, flamby, sir, president, captain, bravo,
charisma, assistantship, arrogant, goodbye, 
strong, debate, failed, incompetent.} 
}; \\
		\node[citizen] (c2){\textbf{Hollande vs Sarkozy (economic, crises, measures, and copy-paste stories).} The central conversations are on Hollande's and Sarkozy's walls. There are many copy-paste comments, such as a derogatory riddle about Hollande and media questions denouncing Sarkozy's corruption. Compare to cluster 1 (above), there are also more detailed themes like economic, crises, and measures taken by politicians.\\  
\textit{Keywords: residential, descent, child, clinical, chic, aristocrat, inhabit, land, employment.} 
}; \&;
		\& \node[post] (p2){\textbf{Fights among multiple candidate's supporters.} The central conversations are on many candidates' walls (Hollande, Bayrou, Dupont-Aignan, Joly, and Lepen), where supporters praise their candidate or denounce others. The copy-paste derogatory riddle about Hollande also appears repeatedly in the central conversations.\\ 
\textit{Keywords: employment, euro, child, residential, pedigree, chic, clinic, pent, inhabit, land.} 
}; \\
		
		\node[citizen] (c3) {\textbf{Islam, religion, and immigration.} The central conversations are mostly on Le Pen's wall, and then Dupont-Aignan's, M\'elenchon's, Hollande's, and Sarkozy's walls, which are mainly about Islam, religion, and immigration.  
        \\ 
\textit{Keywords: 
Koran, Allah, religion, Islam, angel, pig, pork, Muslim, Arab, racist.}
}; \&;
		\& \node[post] (p3) {\textbf{Islam, religion, and immigration.} The central conversations are on Le Pen's wall, and are more coherent on Islam, religion, and immigration compared to $h = 0.035$. \\
\textit{Keywords: Koran, religion, Allah, angel, Islam, pig, Muslim, mosque, Arab, Lyon, racist, church.}
}; \\
		
		\node[citizen] (c4) {\textbf{Pro-M\'elenchon.}  The central conversations are on M\'elenchon's wall and are mostly positive towards him.  
        \\
\textit{Keywords: JLM, comrade, resistance, FDG, front, liberal, revolutionary, human, capital, ecologic.}
}; \&;
		\& \node[post] (p4) {\textbf{Pro-M\'elenchon.} The central conversations are on M\'elenchon's wall and are mostly positive towards him.\\ 
\textit{Keywords: 
JLM, comrade, resistance, FDG, troll, revolutionary, front, human, struggle, liberal, fight.}
}; \\
	};

	\path[draw, line width = .6pt] (c4.east) -- (p4.west);
	\path[draw, line width = 1.7pt] (c3.east) -- (p3.west);
	\path[draw, line width = .4pt] (c1.east) -- (p1.west);

	\end{tikzpicture}
\end{center}
\caption{\textbf{Cluster patterns when $\bm{h = \infty}$}}
\label{fig:diagram3}
\end{figure}
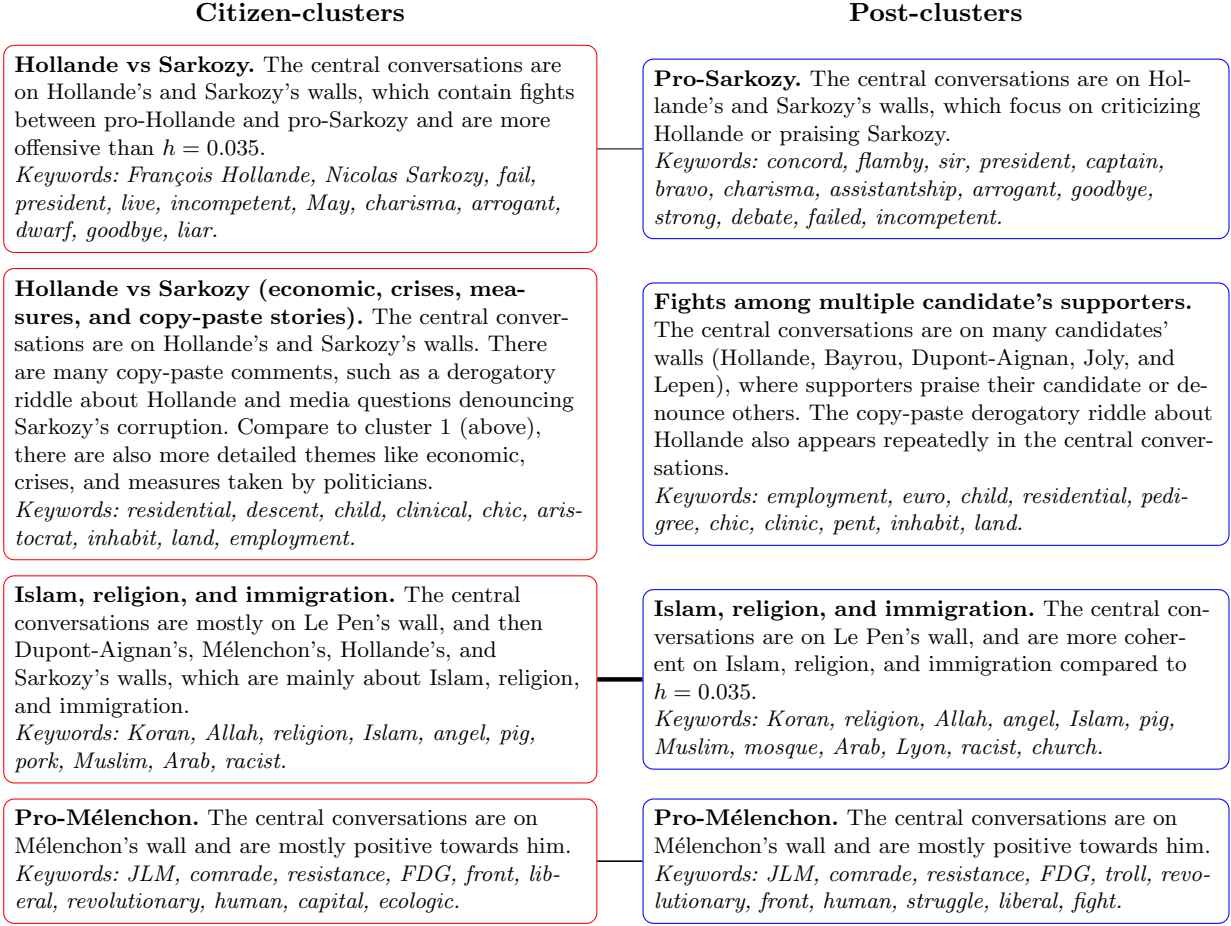
\section{Statistical consistency of \texttt{pairGraphText}}\label{sec:consistency}

This section shows that our graph contextualization method, \texttt{pairGraphText}, is statistically consistent under the Node Contextualized Stochastic co-Blockmodel (NC-ScBM), which is a fusion of the NC-SBM (\cite{binkiewicz2017covariate}) and ScBM (\cite{rohe2016co}).  

\begin{definition}
Let $Z_C\in\{0,1\}^{N_C\times K_C}$ and $Z_P\in\{0,1\}^{N_P\times K_P}$, such that there is only one 1 in each row and at least one 1 in each column.  Let $B\in[0,1]^{K_C\times K_P}$ be of rank $K = \min\{K_C, K_P\}$.  Let $E_C\in \mathbbm{R}^{K_C\times M_C}$ and $E_P\in \mathbbm{R}^{K_P\times M_P}$.  Under the NC-ScBM, the adjacency matrix $A\in\{0,1\}^{N_C\times N_P}$ contains independent Bernoulli random variables with
$$(1)\quad \mathcal{A} = \mathbbm{E}[A] = Z_CBZ_P,$$
and the node covariate matrices $X\in\mathbbm{R}^{N_C\times M_C}$ and $Y\in\mathbbm{R}^{N_P\times M_P}$ contain independent sub-gaussian elements with
$$(2)\quad \mathcal{X} = \mathbbm{E}[X]= Z_CE_C \text{ and } \mathcal{Y} = \mathbbm{E}[Y] = Z_PE_P.$$
\end{definition} 

Recall the similarity matrix for \texttt{pairGraphText} defined in Equation \eqref{def:L}, $S = L+hXT_{\omega}(W)Y^T$.  We define the population similarity matrix as 
\begin{equation}\label{def:Lpop}
\mathcal{S} = \mathcal{L}+h\mathcal{X}\mathcal{W}\mathcal{Y}^T,
\end{equation} where $\mathcal{L}=\mathcal{D}_C^{-1/2}\mathcal{A}\mathcal{D}_P^{-1/2}$ and $\mathcal{W} = \mathcal{X}^T\mathcal{L}\mathcal{Y}$, where diagonal matrices $
[\mathcal{D}_C]_{ii} = \sum_{j}\mathcal{A}_{ij}+\tau_c$
 and 
$[\mathcal{D}_P]_{jj} = \sum_{i}\mathcal{A}_{ij}+\tau_p$.  Let $U_C$ and $\mathcal{U}_{C}\in\mathbbm{R}^{N_C\times K}$($U_P$ and $\mathcal{U}_{P}\in\mathbbm{R}^{N_P\times K}$) contain the top $K$ left(right) singular vectors of $S$ and $\mathcal{S}$. 

The basic outline of the proof for statistical consistency is: Under some conditions,
\begin{enumerate}
\item the element-wise difference between $T_{\omega}(W)$ and $\mathcal{W}$ is bounded by $\omega$ in probability;
\item the similarity matrix $S$ converges to $\mathcal{S}$ in probability; 
\item the singular vectors $U_C$ and $U_P$ converge to $\mathcal{U}_C$ and $\mathcal{U}_P$ within some rotations in probability;  
\item the mis-clustering rates for citizens and candidate-posts goes to zero in probability. 
\end{enumerate}
The definition of mis-clustered is the same as in \cite{rohe2016co} and is given in Section 3.2 in supplementary material.  The complete proof is given in Section 3.3 in supplementary material.

Denote $\|\cdotp\|$ as the spectral norm and $\|\cdotp\|_F$ as the Frobenius norm.  For any matrix $H$, we define 
$sym(H) = \left(\begin{matrix} 
0 &H\\
H^T &0
\end{matrix}\right)$ and $\|H\|_2 = \max(\|\max\limits_{i}\|H_{i\cdotp}\|_2, \max\limits_{j}\|H_{\cdotp j}\|_2)$.  Denote $\|\cdotp\|_{\phi_2}$ as the sub-gaussian norm, such that for any random variable $\xi$, there is $\|\xi\|_{\phi_2}=\sup\limits_{t\geq1}t^{-1/2}(\mathbb{E}|\xi|^t)^{1/t}$.  To simplify notation, we denote $N$ as the number of nodes and $M$ as the number of covariates, though $N_C$ and $N_P$, $M_C$ and $M_P$ can be different.


\begin{theorem}\label{thm:upper}
Suppose $A$, $X$ and $Y$, are the adjacency matrix and the node covariate matrices sampled from the NC-ScBM.   Let $\lambda_1 \geq \lambda_2\geq\cdots\lambda_K>0$ be the $K$ non-zero singular values of $\mathcal{S}$.  Let $\mathcal{M}_C$ and $\mathcal{M}_P$ be the mis-clustered citizens and the mis-clustered candidate-posts.  Denote $q_c$ and $q_p$ as the largest sizes of citizen-clusters and post-clusters.  Define $\delta = \min(\min_i[\mathcal{D}_C]_{ii}, \min_j[\mathcal{D}_P]_{jj})$ and $\gamma = \max(\|X\|_2, \|Y\|_2, \|\mathcal{X}\|_2, \|\mathcal{Y}\|_2)$.  Define 
$\xi = \max(\sigma^2\|L\|_F\sqrt{\ln M}, \sigma^2\|L\|\ln M, \frac{\gamma^2}{\delta}\sqrt{\ln M})$, where $L$ is the regularized graph Laplacian defined in Equation \eqref{def:graphLap} and  $\sigma = \max(\max_{ij}\|X_{ij}-\mathcal{X}_{ij}\|_{\phi_2}, \max_{ij}\|Y_{ij}-\mathcal{Y}_{ij}\|_{\phi_2})$.  For any $\epsilon\in(0,1)$, assume
\begin{align*}
&(1)\quad
\delta>3\ln(2N)+3\ln(8/\epsilon),\\
&(2)\quad \xi = o(\omega),\text{ and }\\
&(3) \quad h\leq
\min(
\frac{a}{\gamma^2\|sym(\mathcal{W})\|},
\frac{a}{\gamma^2\omega})\text{, where }a = \sqrt{\frac{3\ln(16N/\epsilon)}{\delta}}.
\end{align*}
Then, with probability at least $1-\epsilon$, for large enough $N$, the mis-clustering rates 
\begin{align*}
\frac{|\mathcal{M}_C|}{N}\leq\frac{c_0q_cK\ln(16N/\epsilon)}{N\lambda_K^2\delta} \text{ and }
\frac{|\mathcal{M}_P|}{N}\leq\frac{c_0q_pK\ln(16N/\epsilon)}{N\lambda_K^2\delta},
\end{align*}
for some constant $c_0$.
\end{theorem}

\paragraph{Remark}
Assumption (1) indicates the sparsity of the graph.  Assumption (2) and (3) are conditions on parameters $\omega$ and $h$ for consistency.  Note the largest sizes of clusters $q_c$ and $q_p$ are $O(N)$. Suppose $\lambda_K$ is lower bounded by some constant $c_1>0$, which indicates the ``signal" of each of the $K$ blocks is strong enough to be detected.  Then, when $\delta$ grows faster than $\ln N$, we have mis-clustering rates goes to zero as $N\rightarrow\infty$.  



\section{Comparison analysis}\label{sec: compare}


Section \ref{sec: compare_different_w} shows the importance of the call-response matrix $W$.  Section \ref{sec: compare_different_scale} discusses different scaling and weighting choices for document-term matrices.  In Section \ref{sec: compare_rtm}, we compare \texttt{pairGraphText} with state-of-the-art topic modeling approach, relational topic model (RTM) \citep{chang2009relational}, on both the Facebook discussion threads (Section \ref{sec: compare_rtm_facebook}) and on the simulated data (Section \ref{sec: compare_rtm_simulation}).

\subsection{Importance of the call-response matrix $W$}\label{sec: compare_different_w}   Recall the call-response matrix $W$ (defined in \eqref{def:w}), which shows the correlation between citizen-words and thread-words on the communication network.  It induces weights on different pairs of citizen-words and thread-words; word-pairs with higher correlation on the network are weighted more.  In this section, we show the importance of the weights induced by the matrix $W$.  We compare \texttt{pairGraphText} with the \texttt{all-one pairGraphText}, which replaces matrix $W$ by the ``all-one" matrix, $J\in\mathbbm{R}^{M_C\times M_P}$, where $J_{ij} = 1$, for all $i,j$.  For comparison, we set the tunning parameter in \eqref{def:L} as $h = \infty$.  Table \ref{tab: specon} and \ref{tab: now} show the keywords of each cluster by \texttt{pairGraphText} and by \texttt{all-one pairGraphText}.

\begin{table}[!htb]
	\caption{\textbf{Keywords in clusters by \texttt{pairGraphText}}}
	\label{tab: specon}
	\begin{minipage}{.5\linewidth}
		\begin{center}
			\begin{tabular}{|c|c|}
				\hline
				\multicolumn{2}{|c|}{Citizen-Clusters} \\
				\cline{1-2}
				\multirow{3}{*}{Cluster 1}  
				& Fran\c cois Hollande, Nicolas Sarkozy, \\
				& fail, president, live, incompetent, May, \\
				& charisma, arrogant, dwarf, goodbye, liar\\
				\hline
				\multirow{3}{*}{Cluster 2}  
				& residential, descent, child, clinical, chic, \\
				& aristocrat, inhabit, land, employment \\
				& \\
				\hline
				\multirow{3}{*}{Cluster 3}  
				& Koran, Allah, religion, Islam, angel, \\
				& pig, pork, Muslim, Arab, racist\\
				& \\
				\hline
				\multirow{3}{*}{Cluster 4}  
				& JLM, comrade, resistance, FDG, \\
				& front, Jean-Luc M\'elenchon, liberal,\\
				&  revolutionary, human, capital, ecologic \\
				\hline
			\end{tabular}
		\end{center}
	\end{minipage}%
		\begin{minipage}{.5\linewidth}
		\begin{center}
			\begin{tabular}{|c|c|}
				\hline
				\multicolumn{2}{|c|}{Post-Clusters} \\
				\cline{1-2}
				\multirow{3}{*}{Cluster 1}  
				& concord, flamby, sir, president, captain, bravo,\\  
				& charisma, assistantship, arrogant, goodbye, \\
				& strong, debate, failed, incompetent\\
				\hline
				\multirow{3}{*}{Cluster 2}  
			    & employment, euro, child, residential, \\ 
			    & pedigree, chic, clinic, pent, inhabit, land\\
			    & \\
				\hline
				\multirow{3}{*}{Cluster 3}  
				& Koran, religion, Allah, angel, Islam, pig,  \\
				& Muslim, mosque, Arab, Lyon, racist, church \\
				& \\
				\hline
				\multirow{3}{*}{Cluster 4}  
				& JLM, comrade, resistance, FDG, troll, \\
				& Jean-Luc M\'elenchon, revolutionary, front,\\ 
				& human, struggle, liberal, fight\\
				\hline
			\end{tabular}
		\end{center}
	\end{minipage}%
\end{table}

\begin{table}[!htb]
	\caption{\textbf{Keywords in clusters by \texttt{all-one pairGraphText}}}
	\label{tab: now}
	\begin{minipage}{.5\linewidth}
		\begin{center}
			\begin{tabular}{|c|c|}
				\hline
				\multicolumn{2}{|c|}{Citizen-Clusters} \\
				\cline{1-2}
				\multirow{3}{*}{Cluster 1}  
				& identity, trick, fascist, opposite, reducer, \\
				&  Allah, flamby, top, continuous, \\ 
				& incarnate, commercial, mission \\
				\hline
				\multirow{3}{*}{Cluster 2}  
				& baptist, professor, suburb, king, \\
				& happiness,aristocrat, sincerity, school, \\
				& regime, residential, exist, erasable, place \\
				\hline
				\multirow{3}{*}{Cluster 3}  
				& dismissal, fraud, multiple, lump, \\
				& aggravating, unfair, review, gift, \\
				& parliamentary, budget, referendum\\
				\hline
				\multirow{3}{*}{Cluster 4}  
				& Parisian, Russian, discriminant, \\
				& defense, land, vineyard, flag, revel,\\ 
				& pedigree, captain, conceivable, \\
				\hline
			\end{tabular}
		\end{center}
	\end{minipage}%
	\begin{minipage}{.5\linewidth}
		\begin{center}
			\begin{tabular}{|c|c|}
				\hline
				\multicolumn{2}{|c|}{Post-Clusters} \\
				\cline{1-2}
				\multirow{3}{*}{Cluster 1}  
			    & continued, resistance, great, passion, \\
			    & channel, bravo, debate, fight, difficult, \\
			    & goodbye, great, beat, stand, hope\\
				\hline
				\multirow{3}{*}{Cluster 2}  
				& troll, military, comrade, raid, concord, max, \\
				& Philippe Poutou, tomorrow, soldier, killer, \\
				& victim, hateful, bulletin, Jewish, fraud \\
				\hline
				\multirow{3}{*}{Cluster 3}  
				& Allah, Israel, altarpiece, foul, angel, \\
				& dozen, list, cuckoo, municipal\\
				& \\
				\hline
				\multirow{3}{*}{Cluster 4}  
				& lucid, Allah, African, boat, clandestine, \\
				& sister, successful, realist, old, movie, \\
				& angel, tear, promise\\
				\hline
			\end{tabular}
		\end{center}
	\end{minipage}%
\end{table}

Without weights on the word-paris, the \texttt{all-one pairGraphText} fails to extract some topics in the citizen-clusters, such as Islam and the debates among top candidates, which are clear in the citizen-clusters by \texttt{pairGraphText}.  Moreover, some words appear in multiple clusters by the \texttt{all-one pairGraphText}.  For example, the word \textit{Allah} appears in both post-cluster 3 and 4 in Table \ref{tab: now}.  This makes it harder to distinguish different topics between different clusters.  

\subsection{Different choices for document-term matrices}\label{sec: compare_different_scale}
Recall the document-term matrices $X$ and $Y$ (defined in Section \ref{bow_mat}).  These matrices don't consider lengths of comments and posts or popularities of words.  We can address this issue by either (1) scaling the document-term matrices by rows and columns as in Section \ref{issue_structure}, i.e. replacing $X_{ij}$ and $Y_{ij}$ by $$X_{ij}/\sqrt{\sum_i X_{ij}\sum_j X_{ij}}\text{ and }Y_{ij}/\sqrt{\sum_i Y_{ij}\sum_j Y_{ij}}\text{,}$$ or (2) using the weighted document-term matrices.  One standard weighting method is TF-IDF (term frequency-inverse document frequency), which is commonly used in information retrieval and text mining (\cite{salton1975vector}; \cite{joachims1996probabilistic}; \cite{sivic2003video}; \cite{ramos2003using}).  TF-IDF weights words based on both the document length and the word popularity.  For each word $i$ and document $j$, the TF-IDF is 
$$\frac{\text{\# of occurences of word $i$ in document $j$}}{\text{\# of words in document $j$}}\times\log_2\frac{\text{\# of documents}}{\text{\# of documents that contain word $i$}}.$$
In our data, documents are the posts and comments in the discussion threads.  For the weighted document-term matrix of citizens $Y$, we first calculate the TF-IDF matrix of comments, and then add up those comments from the same citizen.  The weighted document-term matrix $X$ is the TF-IDF matrix of posts.   
We compare plain (unscaled and unweighted), scaled, and TF-IDF weighted document-term matrices on the Facebook discussion threads in Section 3 in the supplementary material.

\subsection{Comparison with relational topic model}\label{sec: compare_rtm}

This section compares \texttt{pairGraphText} and Relational Topic Model(RTM) on both Facebook discussion threads (Section \ref{sec: compare_rtm_facebook}) and simulated data (Section \ref{sec: compare_rtm_simulation}).

\subsubsection{Comparison with Relational Topic Model on the Facebook Discussion Threads}\label{sec: compare_rtm_facebook}

 Relational topic model (RTM) \citep{chang2009relational} is a popular approach to extract topics from documents with a network structure (e.g. citation network).  RTM is designed for uni-partite networks, where there is only one type of nodes.  To apply RTM on the bi-partite network with both candidate-posts and citizens, we consider two approaches, (1) symmetrized network (Table \ref{tab: symmetrized}) and (2) co-occurrence network (Table \ref{tab: co-occurrence}).

We define the symmetrized network as a network with posts and citizens, disregarding the different types of nodes.  Recall the adjacency matrix $A\in\mathbbm{R}^{92,226\times 3239}$(defined in \ref{eq:adjMat}), the adjacency matrix for the symmetrized network is $sym(A) = \left(\begin{matrix} 
0 &A\\
A^T &0
\end{matrix}\right)$.
Table \ref{tab: symmetrized} shows the keywords of the four topics by RTM on the symmetrized network.  Words such as \textit{Nicolas Sarkozy} appears in both Cluster 1 and 3, making it hard to distinguish between different topics.  The racial topic (Islam, religion, and immigration), which is clear by \texttt{pairGraphText}, is not that clear in Table \ref{tab: symmetrized}.

We define the co-occurrence network of posts as a network of posts, where the link width is large when the two posts share many citizens who comment frequently on both posts.  Similarly, we define the co-occurrence network of citizens as a network of citizens, where the link width is large when the two citizens comment a lot on many same posts.  We define the adjacency matrix of the co-occurrence network for posts as $A^TA$ and the adjacency matrix of the co-occurrence network for citizens as $AA^T$.  

Table \ref{tab: co-occurrence} shows the keywords of the four topics by RTM on the co-occurrence networks.  Similarly to \texttt{pairGraphText}, RTM also extracts topics like Islam and religion, debates among top candidates, and economic issues.  

\begin{table}[!htb]
	\caption{\textbf{Keywords in topics by RTM on symmetrized network }}
	\label{tab: symmetrized}
	\begin{center}
		\begin{tabular}{|c|c|}
			\hline
			\multicolumn{2}{|c|}{Topics} \\
			\cline{1-2}
			\multirow{3}{*}{Cluster 1}  
			&Nicolas Sarkozy, president, live, \\
			& Fran\c cois Hollande, bravo, France\\
			& courage, all, good, strong, debate \\    
			\hline
			\multirow{3}{*}{Cluster 2}  
			& residential, land, pent, pedigree, clinical,\\
			& inhabit, chic, functional, school,  \\
			& conceivable, childhood \\
			\hline
			\multirow{3}{*}{Cluster 3}  
			& Nicolas Sarkozy, fair, good, other, \\
			& must, polish, can, say, nothing, \\
			& Fran\c cois Bayrou, generation\\
			\hline
			\multirow{3}{*}{Cluster 4}  
			& fair, good, Jean-Luc M\'elenchon, \\
			& nothing, other, speak, share, yes,  \\
			& say, generation, front, racist, Muslim\\
			\hline
		\end{tabular}
	\end{center}
\end{table}

\begin{table}[!htb]
	\caption{\textbf{Keywords in topics by RTM on co-occurrence networks}}
	\label{tab: co-occurrence}
	\begin{minipage}{.5\linewidth}
		\begin{center}
			\begin{tabular}{|c|c|}
				\hline
				\multicolumn{2}{|c|}{Citizen-Topics} \\
				\cline{1-2}
				\multirow{3}{*}{Cluster 1}  
				& Fran\c cois Hollande, Nicolas Sarkozy, \\
				& France, Jean-Luc M\'elenchon, good, \\
				& all, fair, nothing, must, president\\
				\hline
				\multirow{3}{*}{Cluster 2}  
				& residential, clinic, stock market, live, \\
				& childhood, build, school, free, assembly, \\ 
				& departure, functional, bourgeois, depend\\
				\hline
				\multirow{3}{*}{Cluster 3}  
				& Muslim, religion, Islam, speak, racist, \\
				& insult,  Arab, Koran, evil, fear, angel\\
				& \\
				\hline
				\multirow{3}{*}{Cluster 4}  
				& European, financial, public, undertaken, \\
				& billion, public, budget, advice,  bank, \\
				& balance sheet, service, euros, jobs\\
				\hline
			\end{tabular}
		\end{center}
	\end{minipage}%
	\begin{minipage}{.5\linewidth}
		\begin{center}
			\begin{tabular}{|c|c|}
				\hline
				\multicolumn{2}{|c|}{Post-Topics} \\
				\cline{1-2}
				\multirow{3}{*}{Cluster 1}  
			    & Nicolas Sarkozy, Fran\c cois Hollande, \\
			    & social, debate, may, president, victory, \\
			    & change, rich, augment, tax, poor, million\\
				\hline
				\multirow{3}{*}{Cluster 2}  
				& president, franc, sir, live, bravo, all, aim, \\
				& pay, win, good, want, FDG\\
				&\\
				\hline
				\multirow{3}{*}{Cluster 3}  
				& Islam, religion, racist, evil, immigrant, Arab, \\
				& insult, Koran, know, from, Jewish, racism\\
				& \\
				\hline
				\multirow{3}{*}{Cluster 4}  
				& more, good, France, Jean-Luc M\'elenchon,\\
				& all, fair, Nicolas Sarkozy, Fran\c cois Hollande,\\ 
				& speak, can, politic, generation\\
				\hline
			\end{tabular}
		\end{center}
	\end{minipage}%
\end{table}

\subsubsection{Comparison with relational topic model on simulated data}\label{sec: compare_rtm_simulation}

In this section, we use simulation examples to compare \texttt{pairGraphText} and RTM based upon both statistical accuracy and computational running time.  

We simulate documents with links and text, then use \texttt{pairGraphText} and RTM to cluster these documents.  There are two sources of data, (1) links between documents, i.e. graph, and (2) text in the documents, i.e. text.  We compare \texttt{pairGraphText} and RTM in three cases: (1) when both the graph and text contain block information (both signals), (2) when only the graph contains block information (graph signals), and (3) when only text contains block information (text signals).  For each of the three cases, we simulate varying levels of signal strength.  (See more details on how we define signals in the next paragraph).  For each signal level, we simulate 100 random data sets. Each data set consists of 1000 documents and 1000 words in total, with around 200 words and 20 links per document.  In this step, we simulate the documents with links and words under a block model with two blocks, each with around 500 documents and around 500 words. (See more details for the block model in the next paragraph).  On each data set, we run \texttt{pairGraphText} and RTM to partition the 1000 documents into two clusters.  For RTM, we define its estimated cluster label for each document $i$ as $\max\limits_k \text{\# of words in document $i$ belongs to block $k$}$.  

We simulate all the adjacency matrices (graphs) and the document-term matrices (text) under a Degree Corrected Stochastic Blockmodel \citep{karrer2011stochastic}.  Denote $z(i)$ as the block label of any document $i$ and $z_{text}(w)$ as the block label of any word $w$.  Under this model, two documents $i$ and $j$ are linked with each other with probability $\theta_i\theta_j B_{z(i)z(j)}$, and document $i$ contains word $w$ with probability $\theta_i\theta_w^{text} B_{z(i)z_{text}(w)}^{text}$, where $\theta_i$, $\theta_j$, $\theta_w^{text}$ are degree parameters. The element $B_{uv}$ shows the expected number of links between blocks $u,v$, and the element $B_{uv}^{text}$ shows the expected number of appearances for words in block $v$ in documents in block $u$. We define $B \propto \left(\begin{matrix} 
0.1&0.1\\
0.1 &0.1
\end{matrix}\right) +  sig_g\left(\begin{matrix} 
1&0\\
0 &1
\end{matrix}\right)$ and $B^{text} \propto \left(\begin{matrix} 
0.1&0.1\\
0.1 &0.1
\end{matrix}\right) + sig_t  \left(\begin{matrix} 
1&0\\
0 &1
\end{matrix}\right)$, where the graph signal $sig_g$ and the text signal $sig_t \in  {\{1e-1.8, 1e-1.6, \dots, 1e3\}}$ separately show graph links and words contain how much block information.

For \texttt{pairGraphText}, we set weight $h$ so that the first singular values of the graph Laplacian $L$ and the text assisted part $hC_T$ are equal.  We choose the threshold $\omega$ to be the 95\% quantile of non-zero $|W_{ij}|$'s.  We set the number of random starts in the k-means steps (Step 6 and 7 in Algorithm \ref{algorithm}) as $10^4$.  For RTM, we use the function \textit{rtm.collapsed.gibbs.sampler} in the R package \textit{lda}. We set the scalar value of the Dirichlet hyperparameter for topic proportions $\alpha = 0.001$, the scalar value of the Dirichlet hyperparamater for topic multinomials $\eta = 0.1$, the numeric of regression coefficients expressing the relationship between each topic and the probability of link $\beta = (0.5, 0.5)$, and the number of sweeps of Gibbs sampling over the entire corpus to make as $num.iterations = 1e4$.  We set $\alpha$ to be small since we aim to cluster each document to one topic instead of multiple topics.  We set the $num.iterations$ large enough so that the likelihood from each document converges.

\begin{figure}[H]
	\subfigure[][]{\label{fig:mis_clusterate}
		\includegraphics[width=1\columnwidth]{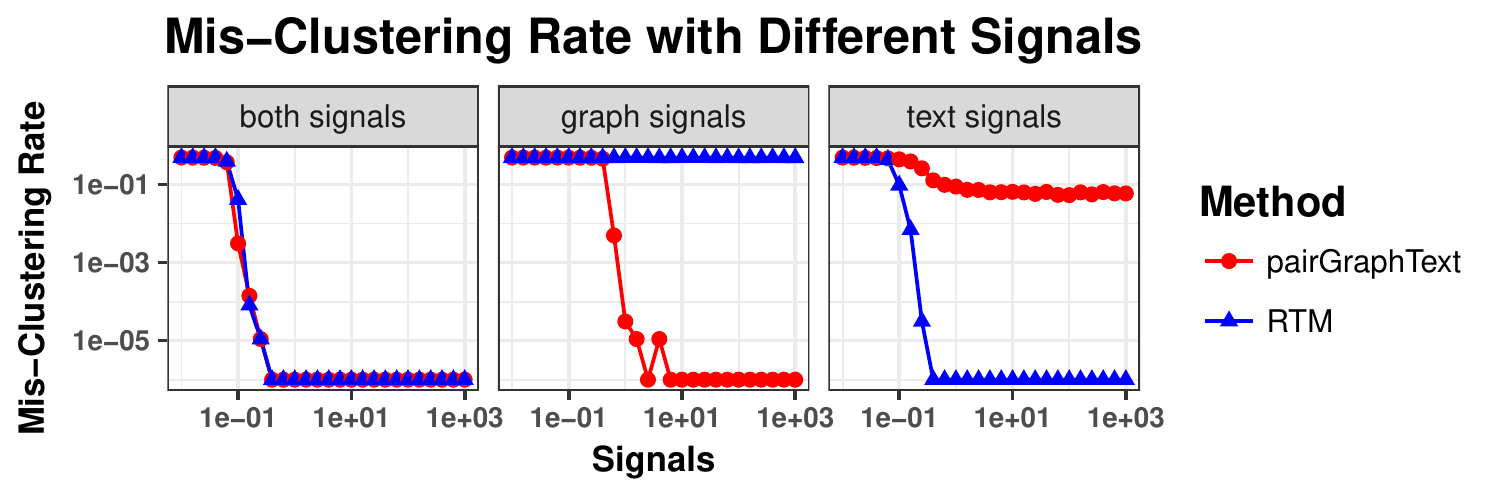}
	}
	\subfigure[][]{\label{fig:runtime}
		\includegraphics[width=1\columnwidth]{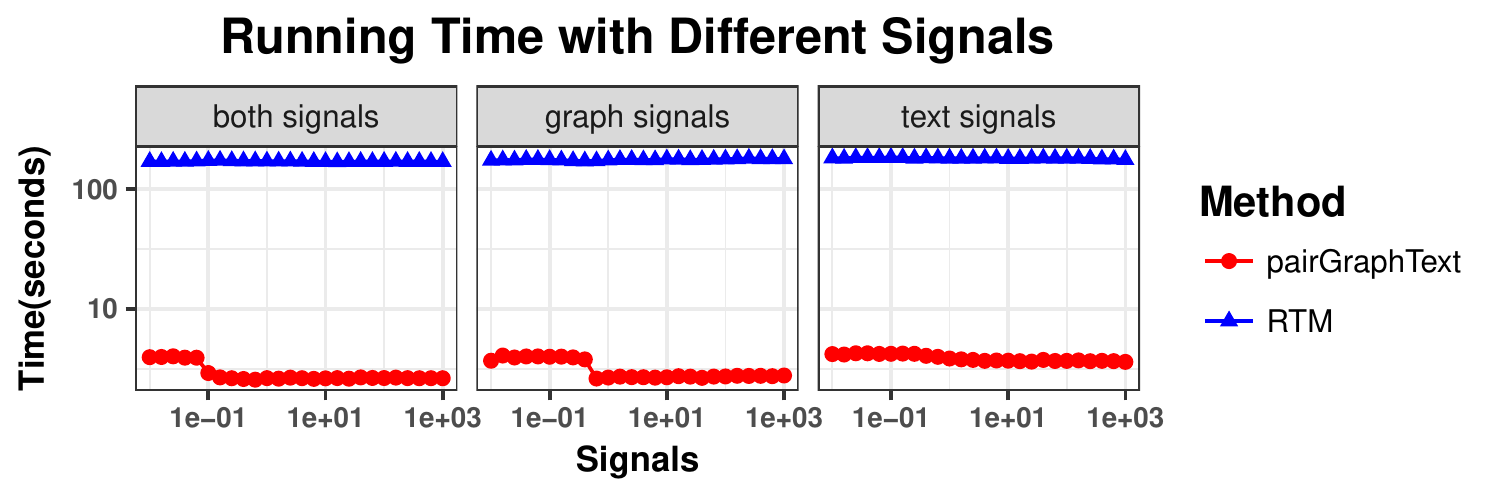}
	}
	
	\caption{\textbf{Comparison between \texttt{pairGraphText} and RTM} }
	\label{fig:rtm_vs_specon}
\end{figure}

Figure \ref{fig:mis_clusterate} compares the mis-clustering rate of \texttt{pairGraphText} and RTM.  Without text signals (the middle plot), RTM fails to recover block labels even with large graph signals, but \texttt{pairGraphText} recovers block labels with the graph signal over 10. Without graph signals (the right plot), \texttt{pairGraphText} can only recover 90\% of block labels, but RTM recovers all block labels, when the text signal is over 0.4.  With both graph signals and text signals (the left plot), both methods perform better than the two cases when only one type of signals exists, and both methods can recover block labels with large enough signals. 

RTM generalizes the text-based topic modeling method, LDA \citep{blei2003latent}, to integrate links (graph); it depends more on text and uses links to improve.  On the other hand, \texttt{pairGraphText} generalizes the link-based spectral clustering to integrate text; it depends more on links and uses text to improve.  From the Figure \ref{fig:mis_clusterate}, RTM fails to recover block labels without text signals, but \texttt{pairGraphText} can still recover most block labels without graph signals.  Figure \ref{fig:runtime} also shows that \texttt{pairGraphText} is much faster than RTM. 

RTM enables us to predict keywords and citations for new documents \citep{chang2009relational}.  However, to cluster massive documents into different topics, \texttt{pairGraphText} is a better choice.

See Section 4 in the supplementary material for more simulations comparing \texttt{pairGraphText} with multiple methods including CASC and spectral clustering.

\section{Discussion}\label{discussion}

This paper searches for (i) candidate-centered structure and (ii) issue-centered structure in the political discussions on Facebook surrounding the 2012 French election.  The candidate-centered structure is relatively easy to detect since we have the labels of each post belongs to which candidate.  But the search for issue-centered structure is more challenging, because we have no such labels of citizens or any labels of issues. To identify topics in the discussions, we use both the graph and the text.  \texttt{pairGraphText} synthesizes the graph and the text, and it adresses the noisy and high-dimensional problem for text by thresholding.  Using \texttt{pairGraphText}, we identify topics that attract people's attention, including Islam, religion, immigration, ecology, economy, and crises.  During the interpretation of clusters, we propose the word-content strategy to extract the cluster topics, and our Shiny App \url{https://yilinzhang.shinyapps.io/FrenchElection} plays a signicant role in the interdisciplinary collaboration between statisticians and social scientists.  Our codes and data sets are available on Github \url{https://github.com/yzhang672/AOAS}.  We also provide an R package \texttt{pairGraphText} to implement our method on Github  \url{https://github.com/yzhang672/pairGraphText}.

\cite{chang2010hierarchical} proposed the relational topic model (RTM), a hierarchical probabilistic model for networks with node covariates.  They modeled topic assignments for documents using latent Dirichlet allocation (LDA) (\cite{blei2003latent}). Instead of studying networks of documents or posts, we study the bi-partite network between candidate-posts and citizens. Also, our method is unsupervised, more computationally efficient, and generally more accurate compared with RTM.  RTM enables us to predict keywords and citations for new documents.  However, to cluster documents into different topics, \texttt{pairGraphText} is a better choice than RTM.

\texttt{pairGraphText} is useful for applications outside of discussion threads.  It is applicable to any network with node covariates. \texttt{pairGraphText} enhances the homogeneity of covariates within clusters. This boosts the signal of the clusters and helps with interpretation.

\section*{Acknowledgements}
We thank Jonathan Chang from Physera and David Blei from Columbia University for their advice in implementing the Relational Topic Models.  We thank Emma Krauska and Fan Chen from University of Wisconsin-Madison for their discussions to name \texttt{pairGraphText}.  We thank the Editor, Associate Editor, and reviewer who provide helpful comments on the manuscript.
\begin{supplement}[id=supp]
  \stitle{Supplementary Materials for Discovering Political Topics in Facebook Discussion threads with Graph Contextualization}
  \slink[url]{http://arxiv.org/src/1708.06872/anc/}
  \sdatatype{.pdf}
  \sdescription{ 
  This supplementary consists of three parts.  Part 1 provides more evidence for the candidate-centered structure.  Part 2 explains our choice of the number of clusters $K$ when searching for the issue-centered structure.  Part 3 discusses different choices for document-term matrices.  Part 4 provides more simulations comparing \texttt{pairGraphText} with RTM and other methods including CASC and spectral clustering.  Part 5 provides theoretical justifications for \texttt{pairGraphText}.  
  }
\end{supplement}

\bibliography{foo}
\bibliographystyle{apalike}

\end{document}